\documentclass[twocolumn,aps,prapplied,nofootinbib,superscriptaddress,amsmath,amssymb,floatfix]{revtex4-2}
\usepackage{graphicx}
\usepackage{dcolumn}	
\usepackage{bm}			
\usepackage{amsfonts}
\usepackage{xspace}
\usepackage{color}
\usepackage{epstopdf}
\usepackage{multirow}
\usepackage{amsmath}
\usepackage{float}
\usepackage{hyperref}
\usepackage{cleveref}
\usepackage{amsmath}
\usepackage{mathtools}
\usepackage{enumitem}
\usepackage[section]{placeins}
\usepackage{soul}
\usepackage{url}
\RequirePackage{hyperref}
\hypersetup{
    breaklinks,
    unicode,
    linktoc=all,
    bookmarksnumbered=true,
    bookmarksopen=true,
    colorlinks,
    linkcolor=blue,
    citecolor=blue,
    urlcolor=blue,
    plainpages=false,
    pdfstartview=FitH,
    pdfborder={0 0 0},
    linktocpage
}

\usepackage{soul}

\begin{document}
\title{Tunable atom-cavity interactions with configurable atomic chains}
\author{Xinwei Li}
\affiliation{Graduate School of China Academy of Engineering Physics, Beijing 100193, P. R. China}
\author{Yijia Zhou}
\affiliation{Graduate School of China Academy of Engineering Physics, Beijing 100193, P. R. China}
\author{Hao Zhang}
\email{hzhang@gscaep.ac.cn}
\affiliation{Graduate School of China Academy of Engineering Physics, Beijing 100193, P. R. China}




\begin{abstract}
We investigate atomic chains with different spatial configurations coupled to a ring cavity comprising two counter-propagating traveling modes. We describe the collective atom-light scattering effect with a structure factor of the atomic chain and demonstrate that the interactions between the atoms and the cavity are controlled by the structure factor, resulting in distinctly different collective excitation modes and energy spectrum compared to Fabry-P\'{e}rot cavities. Remarkably, we observe that a cavity dark mode emerges when the atomic spacings are integer multiples of the half-wavelength. The nodes of this standing-wave dark mode align precisely with the atomic positions, enabling intracavity field conversion without free space scattering. By adjusting the configuration of the atomic chain, we realize tunable photon routing and large optical phase shift with almost no photon loss, which can be used to implement versatile building blocks for optical quantum engineering.
\end{abstract}

\maketitle

\section{Introduction}\label{sec:introduction}
Cavity quantum electrodynamics with cold atoms is a powerful platform with promising potential applications in quantum information processing \cite{haroche2006exploring,reiserer2015cavity,reiserer2022colloquium}. A primary objective in this field is to achieve controlled collective atom-light interactions when many emitters are coupled to the same cavity modes~\cite{ritsch2013cold,chang2018colloquium,mivehvar2021cavity}. This collective coupling plays a crucial role in inducing versatile long-range atom-atom interactions~\cite{norcia2018cavity,periwal2021programmable}, realizing optical phase shifts and photon gates~\cite{vaneecloo2022,stolz2022}, and generating entangled atomic ensembles that are useful in quantum metrology~\cite{leroux2010implementation,haas2014entangled,welte2017cavity,hosten2016measurement,pedrozo2020entanglement,greve2022entanglement}. Recently, there has been growing interest in extending cavity-coupled systems beyond atomic ensembles. Experiments have realized spatially ordered arrays of cold atoms coupled to optical cavities, demonstrating collective cavity scattering~\cite{reimann2015cavity}, superresolution of optical fields~\cite{deist2022superresolution}, high-fidelity midcircuit measurement~\cite{deist2022mid}, and deterministic loading of cavity-coupled arrays~\cite{liu2023realization}.

While most analyses have focused on the interactions between atoms and two-mirror Fabry-P\'{e}rot cavities, ring cavities offer markedly distinct physics~\cite{nagorny2003collective,jia2018strongly,ostermann2020unraveling}. In a Fabry-P\'{e}rot cavity, the intracavity field is a standing wave that is spatially fixed, with its placement determined by the boundary conditions of two cavity mirrors. The atom-cavity coupling strength is maximized when atoms are at the antinodes and minimized when they are at the nodes. In contrast, a ring cavity supports two counterpropagating traveling waves, each with an independent dynamical phase determined by the positions of atoms. Atoms can generate any combination of traveling waves and standing waves inside the cavity. The intracavity field patterns can dynamically follow the atoms, freely translating within the cavity, characterized by a U(1) symmetry~\cite{schuster2020supersolid,mivehvar2018driven}. Our objective is to investigate the intriguing interplay between atom-cavity coupling and the structure of the atomic arrays.

In this paper, we consider an atomic chain that can be configured in different spatial structures. The atomic chain is coupled to a ring cavity that contains two degenerate counterpropagating modes. When the cavity is driven by an external pump laser from a single direction, the backscattering of the atomic chain generates the cavity photons in both directions. The intracavity field pattern is completely determined by the atomic chain. The collective atom-cavity interactions can be controlled by varying the spatial arrangement of the atoms, yielding different collective excitation modes and cavity response spectra. Specifically, when the interatomic distances are integer multiples of the half-wavelength, two cavity standing-wave modes can spontaneously emerge. One of these modes is a ``dark mode," where the nodes of the standing wave align precisely with the atoms. This cavity dark mode is decoupled from the atoms, similarly to the electromagnetically induced transparency (EIT). Conversely, the other mode is a ``bright mode," where the antinodes of the standing wave align with the atoms. Driving the atom-induced dark mode, we achieve lossless optical mode conversion between the two cavity counterpropagating traveling waves. We further analyze the cavity output fields and find that within a broad range of experimentally achievable values for the collective atom-cavity cooperativity, photon routing in two cavity output directions with low photon loss and substantial tuning range can easily be realized with an atomic chain containing a few atoms. Moreover, a large optical phase shift of the cavity output can be realized  by tuning the spatial structure of the atomic chain. Our results demonstrate that atom arrays coupled to a ring cavity with two traveling waves exhibit distinctly different physics from conventional single-mode Fabry-P\'{e}rot cavities in tuning the collective interactions between the atoms and the cavities. This offers new capabilities in manipulating the optical fields with atomic chains and realizing versatile building blocks for applications in quantum information processing~\cite{reiserer2015cavity,reiserer2014quantum,shomroni2014all,scheucher2016quantum,reiserer2022colloquium,xia2013all} and optical quantum computing~\cite{o2007optical,kok2007linear}.

The remainder of this paper is organized as follows. In Sec.~\ref{sec:model}, we present the theoretical model and provide the equations of motion. Sec.~\ref{sec:spectrum} focuses on finding the cavity spectrum under the weak drive and tuning the atom-cavity interactions by the structure factor of the atomic chain. In Sec.~\ref{sec:bright_and_dark}, we explore a specific scenario in which the interatomic distances are integer multiples of the half-wavelength, resulting in two distinct standing-wave modes that are either bright or dark with respect to the atoms. In Sec.~\ref{sec:tunable_output}, we use the input-output relation to determine the relative power and the phases of the cavity output fields, proposing a protocol for achieving low-loss tunable photon routing. In Sec.~\ref{sec:experiment}, we discuss the experimental feasibility of the setup and analyze the effect of atoms' spatial uncertainty on photon-routing processes. Finally, we end with a summary of our findings in Sec.~\ref{sec:conclusion}.

\section{System model}\label{sec:model}
\begin{figure}[!htp]
 \includegraphics[width=1.0\columnwidth]{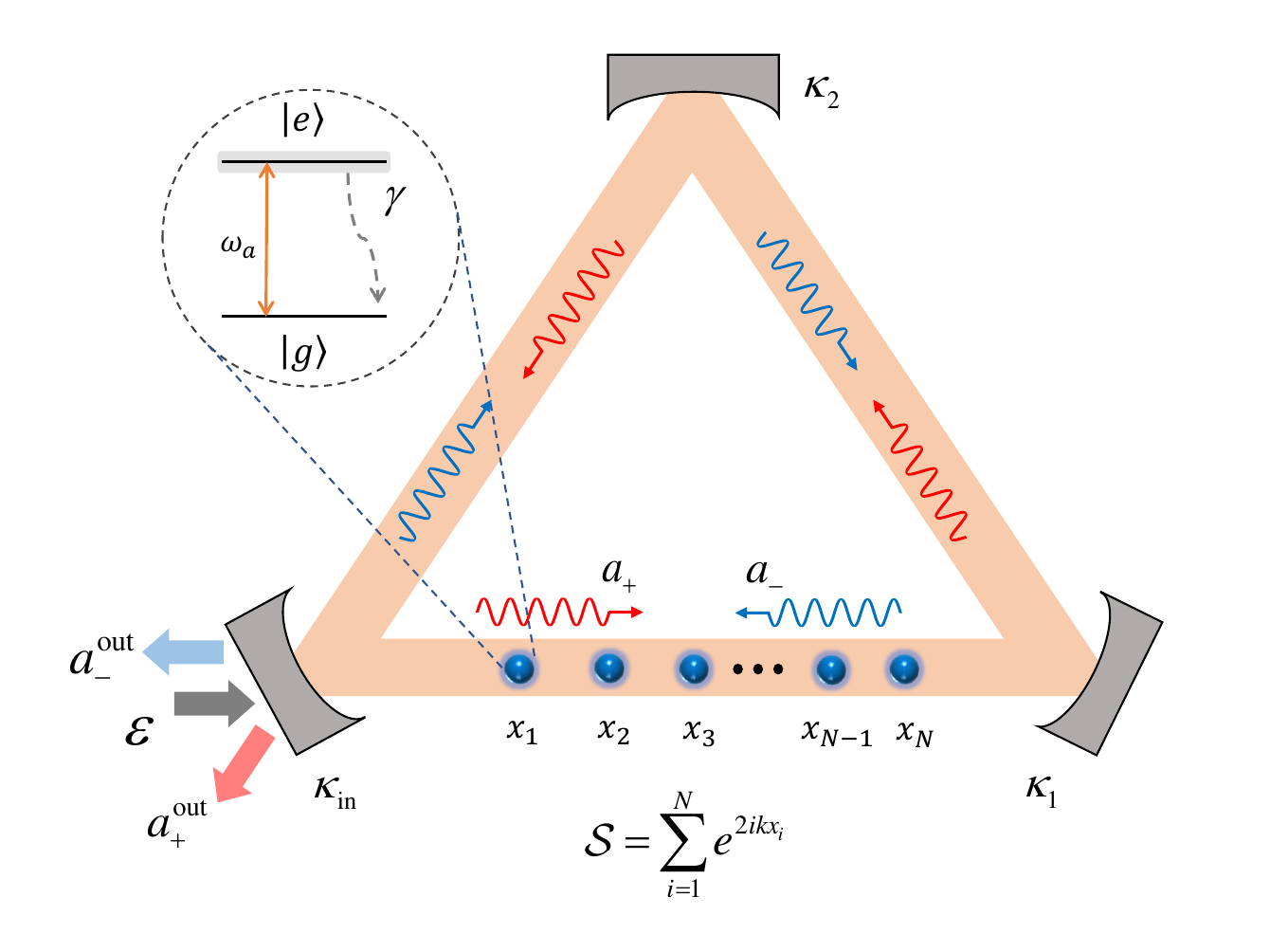}
 \caption{\label{FIG:model} An atomic chain confined within a ring cavity. The system consists of $N$ two-level atoms undergoing internal transitions between the states $|g\rangle \leftrightarrow|e\rangle$. These atoms interact with two counterpropagating cavity modes $\hat{a}_{+}$ and $\hat{a}_{-}$ simultaneously. The position $x_i$ of each atom can be adjusted individually. The configuration of the atomic chain is described by the structure factor $\mathcal{S}$. The decay rate of each cavity mirror is given by $\kappa_i$, while the atomic spontaneous-emission rate is given by $\gamma$. A coherent laser field is injected into the cavity through one of its mirrors to drive the system.}
\end{figure}

As shown in Fig.~\ref{FIG:model}, we consider a chain of $N$ identical two-level atoms trapped in a ring cavity. The ring cavity contains two degenerate counterpropagating modes represented by the annihilation (creation) operators $\hat{a}_{+}$ ($\hat{a}_{+}^{\dagger}$) and $\hat{a}_{-}$ ($\hat{a}_{-}^{\dagger}$) for the forward-propagating and the backward-propagating modes, respectively. The position of each atom, $x_i$, can be adjusted individually. A coherent laser field with frequency $\omega_p$ and amplitude $\varepsilon$ drives the cavity fields. The drive field is injected through one of the cavity mirrors, copropagating with the cavity forward mode. The Hamiltonian in the rotating frame relative to $\omega_p$ is given by
\begin{equation} \label{eq:Ham}
\hat{H}=\hat{H}_0+\hat{H}_I+\hat{H}_P,
\end{equation}
where
\begin{subequations} \label{eq:H}
\begin{align}
\hat{H}_0 & =-\delta \hat{a}_{+}^{\dagger} \hat{a}_{+}-\delta \hat{a}_{-}^{\dagger} \hat{a}_{-}-\Delta \sum_{i=1}^N \hat{\sigma}_i^{+} \hat{\sigma}_i^{-}, \\
\hat{H}_I & =\sum_{i=1}^N\left(g_{i, {-}} \hat{\sigma}_i^{+} \hat{a}_{-}+g_{i, {+}} \hat{\sigma}_i^{+} \hat{a}_{+}\right)+\text {H.c.}, \\
\hat{H}_P & =\sqrt{\kappa_{\mathrm{in}}}\left(\varepsilon^* \hat{a}_{+}+\varepsilon \hat{a}_{+}^{\dagger}\right).
\end{align}
\end{subequations}
Here $\hat{H}_0$ refers to the free Hamiltonian of two cavity modes and atoms, where $\hat{\sigma}_i^{-}=|g\rangle_i\langle e|$ ($\hat{\sigma}_i^{+}=|e\rangle_i\langle g|$) are the lowering (raising) operators of the \emph{i}th atom. Additionally, $\delta=\omega_{p}-\omega_{c}$ and $\Delta=\omega_{p}-\omega_{a}$ denote the detunings of the cavity-mode frequency and the atomic resonance frequencies from the drive laser. $\hat{H}_I$ is the interaction Hamiltonian of the cavity and the atoms under the rotating-wave approximation. It is a straightforward generalization of the Tavis-Cummings model~\cite{tavis1968exact}, with the coupling $g_{i,\nu}$ depending on the propagation directions of the cavity modes and the atomic position, such that $g_{i, {+}}=g e^{i k x_{i}}$ and $g_{i, {-}}=g e^{-i k x_{i}}$, respectively, where $x_i$ denotes the position of the \emph{i}th atom, while $k=\omega_c/c$ is the wave number of the cavity modes. We assume that the cavity modes couple to atomic transitions with the same coupling strengths $\left|g_{i, {-}}\right|=\left|g_{i, {+}}\right|=g$. $\hat{H}_P$ is the Hamiltonian of the drive field, and $\kappa_{\text{in}}$ is the decay rate of the input mirror. Without loss of generality, throughout this work, we consider only the case where $\hat{a}_{+}$ is driven directly.

The dissipations stemming from the atomic spontaneous emission and cavity photon loss are considered within the Lindblad master equation of density matrix $\rho$, which reads ($\hbar=1$)
\begin{equation} \label{eq:master_eq}
\dot{\rho}=-i[\hat{H}, \rho]+\mathcal{L}_a [\rho]+\mathcal{L}_c [\rho],
\end{equation}
where the dissipation operators are
\begin{subequations} \label{eq:Lind}
\begin{align}
\mathcal{L}_a [\rho] &=\sum_{i=1}^N \frac{\gamma}{2}\left(2 \hat{\sigma}_i^{-} \rho \hat{\sigma}_i^{+}-\hat{\sigma}_i^{+} \hat{\sigma}_i^{-} \rho-\rho \hat{\sigma}_i^{+} \hat{\sigma}_i^{-}\right), \\
\mathcal{L}_{c}[\rho]&=\sum_{\nu}\frac{\kappa}{2}\left(2\hat{a}_{\nu}\rho\hat{a}_{\nu}^{\dagger}-\hat{a}_{\nu}^{\dagger}\hat{a}_{\nu}\rho-\rho\hat{a}_{\nu}^{\dagger}\hat{a}_{\nu}\right),
\end{align}
\end{subequations}
with $\nu \in\{{+}, {-}\}$. The atomic spontaneous-emission rate is $\gamma$ and the total cavity decay rate is $\kappa=\kappa_\text {in}+\kappa_1+\kappa_2$.

The time evolution of the operator's expectation values for the system can be obtained with the master equation above:
\begin{subequations} \label{eq:EOM}
\begin{align}
\partial_t\left\langle\hat{a}_{+}\right\rangle & =\left(i\delta-\frac{\kappa}{2}\right)\langle\hat{a}_{+}\rangle-i\sum_{i=1}^{N}g_{i,+}^{\ast}\left\langle\hat{\sigma}_{i}^{-}\right\rangle-i\sqrt{\kappa_{\mathrm{in}}}{\varepsilon}, \\
\partial_t\left\langle\hat{a}_{-}\right\rangle &=\left(i \delta-\frac{\kappa}{2}\right)\left\langle\hat{a}_{-}\right\rangle-i \sum_{i=1}^N g_{i, -}^*\left\langle\sigma_i^{-}\right\rangle, \\
\partial_t\left\langle\hat{\sigma}_i^{-}\right\rangle &=\left(i \Delta-\frac{\gamma}{2}\right)\left\langle\hat{\sigma}_i^{-}\right\rangle+i\sum_{\nu} g_{i, \nu}\left\langle\hat{\sigma}_i^z \hat{a}_\nu\right\rangle, \\
\partial_t\left\langle\hat{\sigma}_i^z\right\rangle & =-\gamma\left(1+\left\langle\hat{\sigma}_i^z\right\rangle\right)-2i\sum_{\nu} g_{i, \nu}\left\langle\hat{\sigma}_i^{+} \hat{a}_\nu\right\rangle+\text { H.c.}.
\end{align}
\end{subequations}
The steady-state solutions of Eq.~\eqref{eq:EOM} can be obtained by setting $\partial_t\left\langle\hat{a}_{+}\right\rangle=\partial_t\left\langle\hat{a}_{-}\right\rangle=\partial_t\left\langle\hat{\sigma}_i^{-}\right\rangle=\partial_t\left\langle\hat{\sigma}_i^z\right\rangle=0$ for $i=1, \cdots, N$.

\section{Tuning atom-cavity interactions with structure factor}\label{sec:spectrum}
To understand the effect of the atomic chain spatial configurations on the atom-cavity interactions, we examine the system under the drive field $\varepsilon$ by applying the weak-excitation approximation $\langle\hat{\sigma}_i^z\hat{a}_\nu\rangle\approx-\langle\hat{a}_\nu\rangle$. The equations governing the cavity fields are derived by elimination of $\langle\hat{\sigma}_i^-\rangle$ 
\begin{subequations}\label{eq:a_coupled}
\begin{align}
\partial_t\langle \hat{a}_+\rangle &=i\left( \tilde{\delta}-\frac{Ng^2}{\tilde{\Delta}} \right) \langle \hat{a}_+\rangle -\frac{ig^2\mathcal{S}^*}{\tilde{\Delta}}\langle \hat{a}_-\rangle -i\sqrt{\kappa _{\mathrm{in}}}\varepsilon, \\
\partial_t\langle \hat{a}_-\rangle &=i\left( \tilde{\delta}-\frac{Ng^2}{\tilde{\Delta}} \right) \langle \hat{a}_-\rangle -\frac{ig^2\mathcal{S}}{\tilde{\Delta}}\langle \hat{a}_+\rangle,
\end{align}
\end{subequations}
where $\tilde{\Delta}=\Delta+i\gamma/2$, $\tilde{\delta}=\delta+i\kappa/2$. This shows that the two cavity traveling-wave modes are coupled to each other through the backscattering of the atomic chain. Here we define the structure factor
\begin{equation}\label{eq:s}
\mathcal{S}=\sum_{i=1}^N e^{2 i k x_i}
\end{equation}
to describe the effect of the atomic chain configuration on the scattering between the two cavity traveling waves, where the factor of $2 k$ is due to the scattering from the forward mode $\hat{a}_{+}$ to the backward mode $\hat{a}_{-}$. Our structure factor is analogous to the structure factor used in condensed matter physics and crystallography. By adjustment of the axial position of each atom individually, the absolute value $|\mathcal{S}|$ can be varied within the range from 0 to $N$, and the phase of $\mathcal{S}$ is determined by the center of mass of the atom array.

\begin{figure*}[!htp]
 \centering\includegraphics[width=1.0\linewidth]{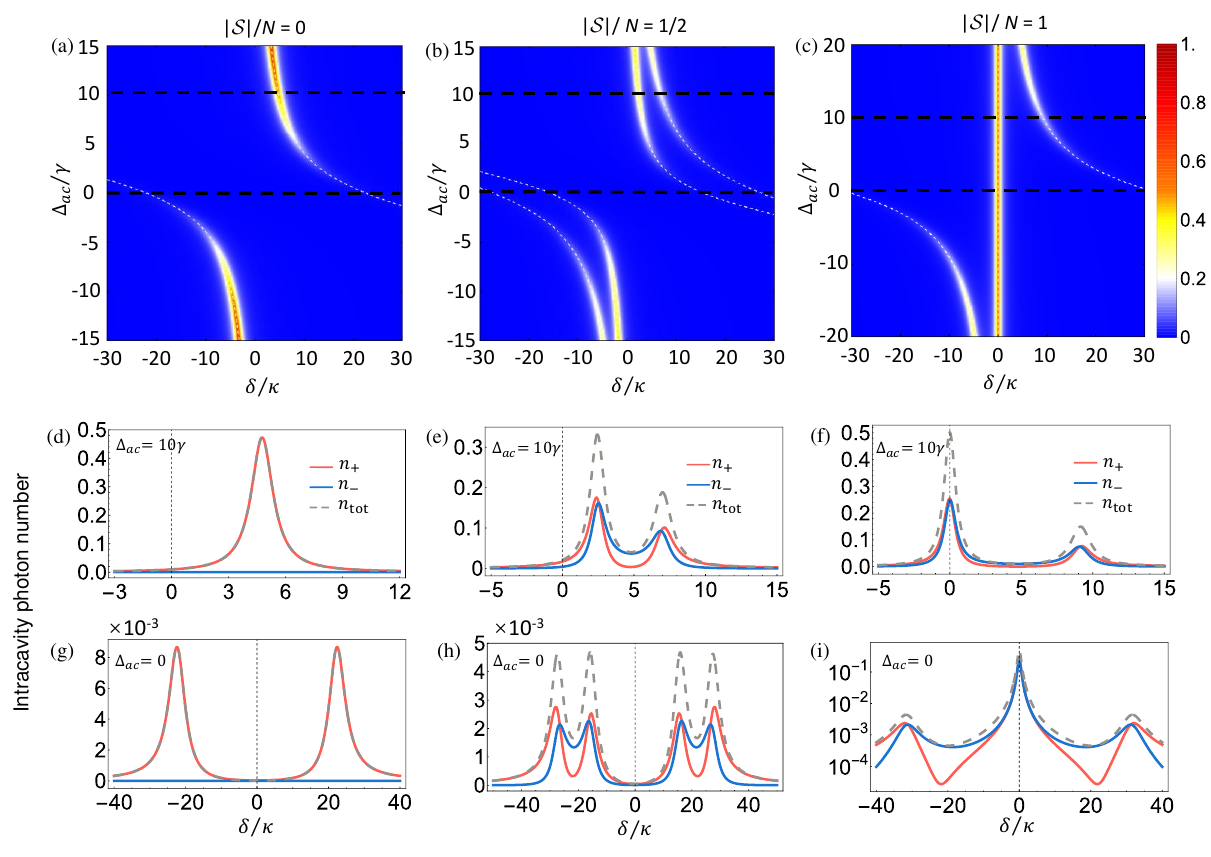}
 \caption{\label{FIG:intra} Mean values of the normalized intracavity photon number for three different atomic spatial configurations described by $|\mathcal{S}|/N$, when the cavity is driven by a pump laser propagating in the direction of the $\hat{a}_{+}$ mode. (a)-(c) Normalized total intracavity photon number $n_{\mathrm{tot}}$ as a function of the atom-cavity detuning $\Delta_{ac}=\omega_c-\omega_a$ and the drive-cavity detuning $\delta=\omega_p-\omega_c$. The eigenenergies of the system, determined by Eq.~(\ref{eq:eigenvalues}), are represented by dashed white lines. The cuts through the dashed black lines are shown in (d)-(i). The normalized intracavity photon numbers $n_{+}$ in the $\hat{a}_{+}$ mode (solid red line), $n_{-}$ in the $\hat{a}_{-}$ mode (solid blue line), and $n_{\mathrm{tot}}$ (dashed gray line) as a function of $\delta$ for cases when the atoms are (d)-(f) detuned from the cavity ($\Delta_{ac}=10\gamma$) and (g)-(i) resonant with the cavity ($\Delta_{ac}=0$). The other parameters are $\kappa=0.1\gamma$, $g=0.5\gamma$, and $N=20$.}
\end{figure*}

Eq.~\eqref{eq:a_coupled} demonstrates that despite the drive field $\varepsilon$ being injected through the input mirror and copropagating with the cavity mode $\hat{a}_+$, the other cavity field $\hat{a}_-$ can be generated as a result of coupling between two cavity modes $\hat{a}_+$ and $\hat{a}_-$ as long as $|\mathcal{S}|\neq0$. To explain this phenomenon, we perform the cavity basis transformations 
\begin{subequations}\label{eq:uncoupled_modes}
\begin{align}
\hat{c}_1&=\frac{1}{\sqrt{2}}\left(\frac{\mathcal{S}}{|\mathcal{S}| }\hat{a}_++\hat{a}_- \right),\\
\hat{c}_2&=\frac{1}{\sqrt{2}}\left(\frac{\mathcal{S}}{|\mathcal{S}| }\hat{a}_+-\hat{a}_- \right),
\end{align}
\end{subequations}
when $|\mathcal{S}|\neq0$. Therefore, Eq.~\eqref{eq:a_coupled} become a new set of equations:
\begin{subequations}\label{eq:a_coupled_trans}
\begin{align}
\partial _t\left< \hat{c}_1 \right> =i\left( \tilde{\delta}-\frac{Ng^2}{\tilde{\Delta}}-\frac{\left| \mathcal{S} \right|g^2}{\tilde{\Delta}} \right) \left< \hat{c}_1 \right> -i\sqrt{\frac{\kappa _{\mathrm{in}}}{2}}\frac{\mathcal{S}}{|\mathcal{S} |}\varepsilon, \\
\partial _t\left< \hat{c}_2 \right> =i\left( \tilde{\delta}-\frac{Ng^2}{\tilde{\Delta}}+\frac{\left| \mathcal{S} \right|g^2}{\tilde{\Delta}} \right) \left< \hat{c}_2 \right> -i\sqrt{\frac{\kappa _{\mathrm{in}}}{2}}\frac{\mathcal{S}}{|\mathcal{S} |}\varepsilon,
\end{align}
\end{subequations}
where the cavity modes $\hat{c}_1$ and $\hat{c}_2$ are decoupled from each other. It is evident from Eq.~\eqref{eq:a_coupled_trans} that the cavity drive field $\varepsilon$ in the $\hat{a}_+$ direction simultaneously excites both the $\hat{c}_1$ mode and the $\hat{c}_2$ mode, with the steady-state intracavity fields obtained by solving Eq.~\eqref{eq:a_coupled_trans} as
\begin{subequations} \label{eq:steady_state_solution_uncouple}
\begin{align}
\left<\hat{c}_{1}\right>_{ss} =\frac{\mathcal{S}\sqrt{\kappa_{\mathrm{in}}}\varepsilon}{\sqrt{2}\left|\mathcal{S}\right|\left(\tilde{\delta}-\frac{Ng^{2}}{\tilde{\Delta}}-\frac{\left|\mathcal{S}\right|g^{2}}{\tilde{\Delta}}\right)}, \\
\left<\hat{c}_{2}\right>_{ss} =\frac{\mathcal{S}\sqrt{\kappa_{\mathrm{in}}}\varepsilon}{\sqrt{2}\left|\mathcal{S}\right|\left(\tilde{\delta}-\frac{Ng^{2}}{\tilde{\Delta}}+\frac{\left|\mathcal{S}\right|g^{2}}{\tilde{\Delta}}\right)}.
\end{align}
\end{subequations}
Transformation of the decoupled basis $\hat{c}_{1}$, $\hat{c}_{2}$ back to the original traveling-wave basis $\hat{a}_{+}$, $\hat{a}_{-}$ with use of Eq.~\eqref{eq:uncoupled_modes} gives the steady-state solution of $\langle\hat{a}_{+}\rangle$ and $\langle\hat{a}_{-}\rangle$ as
\begin{subequations} \label{eq:steady_state_solution}
\begin{align}
\left\langle\hat{a}_{+}\right\rangle_{\text{ss}} &= \frac{\tilde{\Delta}\left(\tilde{\Delta} \cdot \tilde{\delta}-g^2 N\right)  \sqrt{\kappa_{\text {in }}}\varepsilon}{\left(\tilde{\Delta} \cdot \tilde{\delta}-g^2 N\right)^2-g^4|\mathcal{S}|^2},\\
\left\langle\hat{a}_{-}\right\rangle_{\text{ss}} &= \frac{\tilde{\Delta} g^2 \mathcal{S} \sqrt{\kappa_{\text {in }}}\varepsilon} {\left(\tilde{\Delta} \cdot \tilde{\delta}-g^2 N\right)^2-g^4|\mathcal{S}|^2},
\end{align}
\end{subequations}
which can also be derived from Eq.~\eqref{eq:a_coupled} directly.

In Fig.~\ref{FIG:intra}, we present the normalized steady-state intracavity photon number for each cavity mode $n_{+}= \kappa\left|\left\langle\hat{a}_{+}\right\rangle_{\text{ss}}\right|^2/4|\varepsilon|^2$ and $n_{-}= \kappa\left|\left\langle\hat{a}_{-}\right\rangle_{\text{ss}}\right|^2/4|\varepsilon|^2$, as well as the total intracavity photon number $n_{\mathrm{tot}}=n_{+}+n_{-}$ for various values of $|\mathcal{S}|$. Here the normalization factor for $n_{+}$ and $n_{-}$ is the photon number of the resonantly driven empty cavity $4|\varepsilon|^2/\kappa$. For simplicity, we consider that the ring cavity has only one partially transmitting mirror, while all the other mirrors are perfectly reflecting, such that $\kappa_\mathrm{in} = \kappa$. When $|\mathcal{S}|$ is minimized such that $|\mathcal{S}|=0$, the spectrum of the system resembles that of a single-mode Fabry–P\'{e}rot cavity. The $\hat{a}_{-}$ mode is not excited in this case because the backscattering of the atomic chain is zero. For intermediate values of $|\mathcal{S}|$, for example, $|\mathcal{S}|/N=1/2$, the spectrum exhibits two sets of shifted cavity spectra. When the atomic chain is ordered such that the atomic spacings are integer multiples of $\lambda/2$, $|\mathcal{S}|$ reaches its maximum value such that $|\mathcal{S}|/N=1$. In this scenario, the spectrum of $n_{\mathrm{tot}}$ has an unshifted cavity mode centered at $\delta=0$. 

\begin{figure}[!htp]
 \includegraphics[width=0.95\columnwidth]{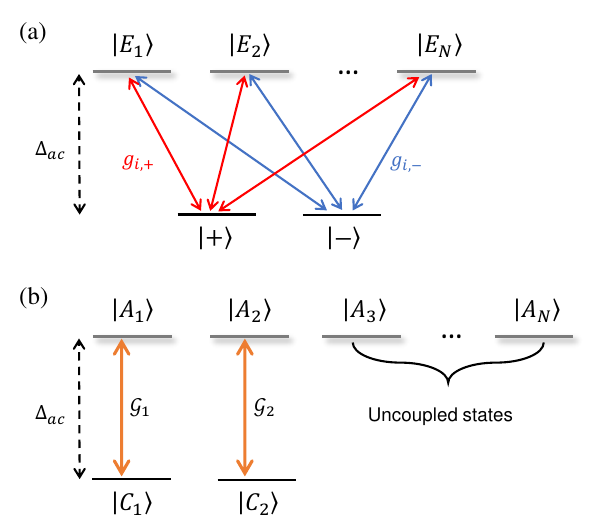}
 \caption{\label{FIG:level} Energy-level diagram of the system. (a) Two counterpropagating cavity traveling modes $|+\rangle$ and $|-\rangle$ are coupled to each of $N$ atoms with position-dependent coupling $g_{i,+}=ge^{ik x_{i}}$ and $g_{i,-}=ge^{-ik x_{i}}$ as in Eq.~(\ref{eq:Ham_single}). (b) The collective interactions between the relevant atomic excitations and the cavity modes are better revealed with two superposed cavity modes $\left|C_1\right\rangle=\frac{1}{\sqrt{2}}\left(\frac{\mathcal{S}^*}{|\mathcal{S}|}|+\rangle+|-\rangle\right)$ and $\left|C_2\right\rangle=\frac{1}{\sqrt{2}}\left(\frac{\mathcal{S}^*}{|\mathcal{S}|}|+\rangle-|-\rangle\right)$ that are coupled to two collective atomic excitation modes $|A_{1}\rangle=\frac{1}{\mathcal{N}}\sum_{i=1}^{N}\alpha_{i}^{+}|E_{i}\rangle$ and $|A_{2}\rangle=\frac{1}{\mathcal{N}}\sum_{i=1}^{N}\alpha_{i}^{-}|E_{i}\rangle$ with coupling $\mathcal{G}_1=g \sqrt{N+|\mathcal{S}|}$ and $\mathcal{G}_2=g \sqrt{N-|\mathcal{S}|}$, respectively, as in Eq.~(\ref{eq:Ham_single_2}). The remaining $N-2$ atomic excitation modes $|A_{3}\rangle$, ... ,$|A_{N}\rangle$ are decoupled from the cavity.}
\end{figure}

In addition to the analysis of intracavity optical fields, we also investigate the collective atomic excitations. We simplify the system within the single-excitation states, assuming that the cavity fields and atoms are weakly excited. This assumption allows us to confine our focus to the following relevant states: the photonic single-excited states, $|{+}\rangle=\hat{a}^{\dagger}_{+}|\mathrm{vac}\rangle$ and $|{-}\rangle=\hat{a}^{\dagger}_{-}|\mathrm{vac}\rangle$, respectively, as well as $\left|E_i\right\rangle=\hat{\sigma}^+_i|\mathrm{vac}\rangle$, where $|\mathrm{vac}\rangle$ is the vacuum state, with the number of cavity photons being zero and all of the atoms being in the ground state. By restricting our analysis to this basis, the effective undriven Hamiltonian is expressed as follows:
\begin{equation}\label{eq:Ham_single}
\begin{aligned} 
\hat{H}_{\mathrm{eff}}&=\tilde{\omega}_c(|+\rangle\langle+|+|-\rangle\langle-|)+\tilde{\omega}_a\sum_{i=1}^N\left|E_i\right\rangle\left\langle E_i\right| \\ 
&+\sum_{i=1}^N\left(g_{i,+}|+\rangle\left\langle E_i\left|+g_{i,-}\right|-\right\rangle\left\langle E_i\right|\right)+\text {H.c.},
\end{aligned}
\end{equation}
where $\tilde{\omega}_c=\omega_c-i\kappa/2$ and $\tilde{\omega}_a=\omega_a-i\gamma/2$. Eq.~\eqref{eq:Ham_single} can also be equivalently represented in matrix form as
\begin{equation}
\left(\begin{array}{c|c}
\mathbf{H}_{\mathrm{cav}} & \mathbf{G} \\
\hline \mathbf{G}^{\dagger} & \mathbf{H}_{\text{atom}}
\end{array}\right). \nonumber
\end{equation}
Here $\mathbf{H}_{\mathrm{cav}}=\tilde{\omega}_{c}\mathbb{I}_2$ represents the cavity modes and $\mathbf{H}_{\mathrm{atom}}=\tilde{\omega}_{a}\mathbb{I}_N$ represents the atoms, where $\mathbb{I}_N$ is the $N$-by-$N$ identity matrix. The interaction terms are as follows:
\begin{equation}
\mathbf{G}=\left(\begin{array}{ccc}
g e^{-i k x_1} & \cdots & g e^{-i k x_N} \\
g e^{i k x_1} & \cdots & g e^{i k x_N}
\end{array}\right). \nonumber
\end{equation}
We decompose $\mathbf{G}$ using singular-value decomposition as $\mathbf{G}=\mathbf{U} \boldsymbol{\Sigma} \mathbf{V}^{\dagger}$~\cite{wickenbrock2013collective,emary2013dark}. When $|\mathcal{S}|>0$, from $\mathbf{U}$ we find two superposed single-photon states of the cavity $\hat{c}_{1}$ and $\hat{c}_{2}$ modes as follows:
\begin{subequations}\label{eq:collective_cav}
\begin{align}
\left|C_1\right\rangle=\frac{1}{\sqrt{2}}\left(\frac{\mathcal{S}^*}{|\mathcal{S}|}|+\rangle+|-\rangle\right),\\
\left|C_2\right\rangle=\frac{1}{\sqrt{2}}\left(\frac{\mathcal{S}^*}{|\mathcal{S}|}|+\rangle-|-\rangle\right).
\end{align}
\end{subequations}
From $\mathbf{V}$ we find two collective atomic excitations
\begin{subequations}\label{eq:collective_atom}
\begin{align}
\left|A_1\right\rangle=\frac{1}{\mathcal{N}} \sum_{i=1}^N \alpha_i^{+}\left|E_i\right\rangle, \\
\left|A_2\right\rangle=\frac{1}{\mathcal{N}} \sum_{i=1}^N \alpha_i^{-}\left|E_i\right\rangle,
\end{align}
\end{subequations}
coupled to the cavity modes $\left|C_1\right\rangle$ and $\left|C_2\right\rangle$, respectively. Here $\alpha_i^{\pm}=e^{i k x_i} \mathcal{S}^*\pm e^{-i k x_i}|\mathcal{S}|$ represents the coefficient of each component and $\mathcal{N}$ is the normalization factor. The nonzero elements of $\boldsymbol{\Sigma}$ are the singular values of $\mathbf{G}$ given by
\begin{subequations}\label{eq:singular}
\begin{align}
\mathcal{G}_1=g \sqrt{N+|\mathcal{S}|}, \\
\mathcal{G}_2=g \sqrt{N-|\mathcal{S}|}.
\end{align}
\end{subequations}
$\mathcal{G}_1$ is the coupling between the cavity mode $\left|C_{1}\right\rangle$ and the collective atomic excitation $\left|A_{1}\right\rangle$, and $\mathcal{G}_2$ is the coupling between $\left|C_{2}\right\rangle$ and $\left|A_{2}\right\rangle$, as shown in Fig.~\ref{FIG:level}(b). Consequently, the Hamiltonian Eq.~\eqref{eq:Ham_single} can be rewritten as
\begin{equation} \label{eq:Ham_single_2}
\hat{H}_{\mathrm{eff}}=\sum_{i=1}^2( {\tilde{\omega}_c}\left|C_i\right\rangle\left\langle C_i\right|+{\tilde{\omega}_a}\left|A_i\right\rangle\left\langle A_i\right|+\mathcal{G}_i\left|C_i\right\rangle\left\langle A_i\right|+\text{H.c.}).
\end{equation}
From this expression, we observe that the system can be essentially interpreted as two cavity modes interacting with two atomic excitations, respectively. The eigenvalues of the system are obtained by diagonalization of the Hamiltonian Eq.~\eqref{eq:Ham_single_2}, which yields
\begin{equation}\label{eq:eigenvalues}
E_{k, \pm}=\frac{(\tilde{\omega}_a+\tilde{\omega}_c) \pm \sqrt{(\tilde{\omega}_a-\tilde{\omega}_c)^2+4 \mathcal{G}_k^2}}{2},
\end{equation}
where $k=1,2$. Eq.~\eqref{eq:eigenvalues} illustrates maximally four nondegenerate polariton states. The real part corresponds to the system's spectrum as depicted by the dashed white lines in Fig.~\ref{FIG:intra}(a-c) and the imaginary part corresponds to the linewidth of the spectral peaks. When the cavity and atoms are on resonance ($\Delta_{ac}=0$), the eigenfrequencies are $\pm g \sqrt{N \pm|\mathcal{S}|}$. In the dispersive regime, the frequency shifts of the cavity modes are $\pm g^2(N \pm|\mathcal{S}|) / \Delta_{ac}$. 

When $|\mathcal{S}|=0$, the singular-value-decomposition process reveals that $\left|C_1\right\rangle=|+\rangle$ and $\left|C_2\right\rangle=|-\rangle$, with the corresponding collective atomic excitations given by $|A_1\rangle =\frac{1}{\sqrt{N}}\sum_{i=1}^N{e^{ik x_i}|E_i\rangle}$ and $|A_2\rangle =\frac{1}{\sqrt{N}}\sum_{i=1}^N{e^{-ik x_i}|E_i\rangle}$ and the couplings given by $\mathcal{G}_1=\mathcal{G}_2=g \sqrt{N}$. This indicates that the $\hat{a}_{+}$ and $\hat{a}_{-}$ traveling modes are independently coupled with the atomic chain. Therefore, the drive field $\varepsilon$ propagating in the forward direction will excite only the cavity $\hat{a}_{+}$ mode and the photon number of the $\hat{a}_{-}$ mode $n_- =0$ as shown in Fig.~\ref{FIG:intra}(d, g). In this special case, the spectrum of the system is reminiscent of the spectrum for a standing-wave Fabry-P\'{e}rot cavity.

\section{Cavity bright and dark modes when $|\mathcal{S}|/N=1$}\label{sec:bright_and_dark}
\begin{figure}[!htp]
 \includegraphics[width=0.95\columnwidth]{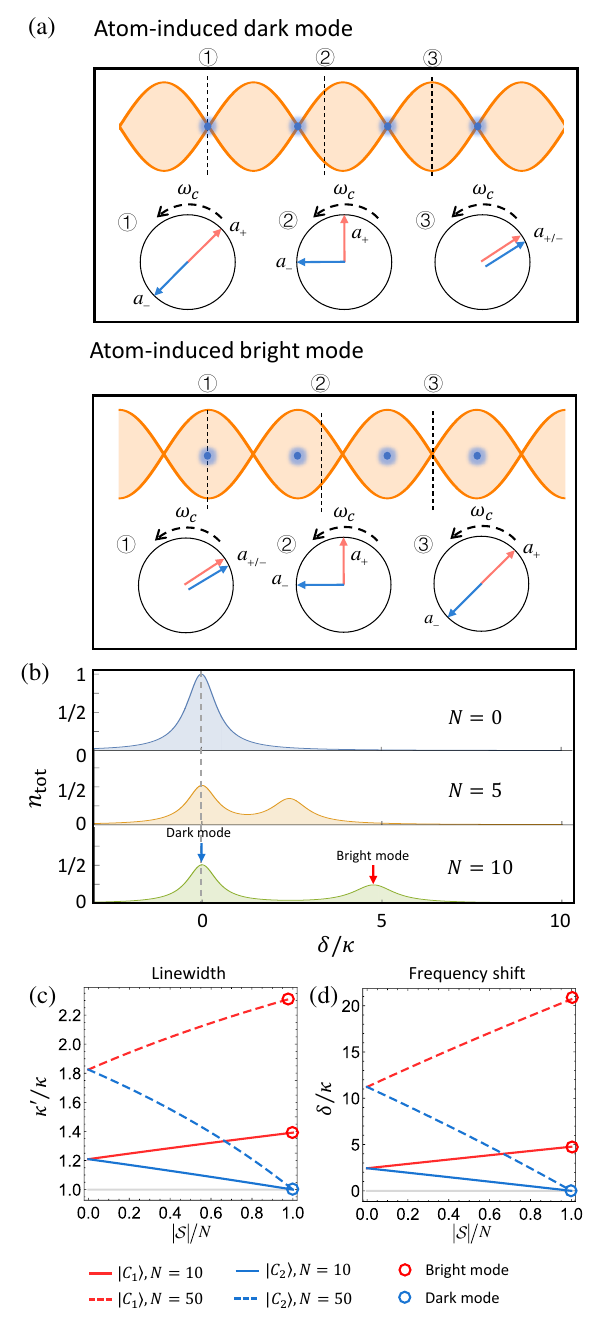}
 \caption{\label{FIG:standing_wave} (a) The dark mode and the bright mode generated by atoms when $|\mathcal{S}|/N=1$. The nodes of the dark mode and the antinodes of the bright mode are aligned with the atoms. The red and blue arrows represent the intracavity fields of the $\hat{a}_{+}$ and $\hat{a}_{-}$ traveling modes respectively. The angle between these two arrows denotes the relative phase between the two traveling waves in different positions. (b) Total normalized intracavity photon number given $|\mathcal{S}|/N=1$ for different values of $N$. The cavity dark mode $\left|C_2\right\rangle$ is centered at $\delta = 0$. The shift and the broadening of the bright mode $\left|C_1\right\rangle$ increase with $N$. (c) Linewidths and (d) frequency shifts  of the $\left|C_1\right\rangle$ (red) and $\left|C_2\right\rangle$ (blue) cavity modes as a function of $|\mathcal{S}|$. When $|\mathcal{S}|/N=1$, $\left|C_1\right\rangle$ and $\left|C_2\right\rangle$ correspond to the cavity bright and dark mode, respectively. The other parameters are $\kappa=0.1\gamma$, $g=0.5\gamma$, $\Delta=10\gamma$, and $N=10$.}
\end{figure}

The atomic chain is maximally structured with $|\mathcal{S}|/N=1$ when the atoms are distributed with the separations equal to integer multiples of the half-wavelength. This configuration gives rise to a cavity dark mode $\left|C_2\right\rangle$ that is decoupled from the atoms since the coupling $\mathcal{G}_2=g\sqrt{N-|\mathcal{S}|}=0$. The other cavity mode $\left|C_1\right\rangle$ is referred to as the bright mode with the collective coupling $\mathcal{G}_1=g \sqrt{N+|\mathcal{S}|}=g\sqrt{2 N}$. The dark mode corresponds to the central peak at $\delta=0$ with the bare cavity linewidth $\kappa$ for any arbitrary $\Delta_{ac}$ as depicted in Fig.~\ref{FIG:intra}(a). Remarkably, cavity modes $\left|C_1\right\rangle$ and $\left|C_2\right\rangle$ are two distinct standing-wave modes that are spontaneously generated through backscattering of the cavity light by the atoms. As shown in Fig.~\ref{FIG:standing_wave}(a), for the atom-induced dark mode $\left|C_2\right\rangle$, the standing wave is formed such that the locations of the nodes are aligned with the atoms. For the bright mode $\left|C_1\right\rangle$, the antinodes are aligned with the atoms. Note that the standing waves of $\left|C_1\right\rangle$ and $\left|C_2\right\rangle$ will dynamically follow the atoms when the atomic chain is displaced.

Figure.~\ref{FIG:standing_wave}(b)-(d) show the frequencies and the linewidths of the bright and dark modes in the dispersive regime. The bright mode $\left|C_1\right\rangle$ is excited when the drive field is tuned at the bright-mode eigenfrequency $\delta \simeq 2 N g^2 / \Delta_{ac}$. In this case, the free-space scattering is twice as much as that in a Fabry-P\'{e}rot cavity, resulting in an increased linewidth $\kappa' \approx \kappa + \frac{2N\gamma}{\Delta_{ac}^2+\gamma^2/4} g^2$. 

Conversely, the dark mode $\left|C_2\right\rangle$ is excited when the drive field is tuned to $\delta=0$. This situation presents a seemingly counterintuitive photon transfer. Even though the drive field $\varepsilon$ directly excites only the $\hat{a}_{+} $ mode, an intracavity field with equal amplitudes of the $\hat{a}_{+}$ and $\hat{a}_{-}$ traveling waves is established by the atoms through backscattering. However, the atoms reside at the nodes of the induced cavity standing wave. Therefore there are no photons scattered into the free space due to the zero coupling strength between the cavity and the atoms at the nodes, resulting in a nonbroadened linewidth $\kappa'=\kappa$ in Fig.~\ref{FIG:standing_wave}(c).

How can atoms transfer photons from the $\hat{a}_{+} $ mode to the $\hat{a}_{-} $ mode while residing at the nodes of the standing wave and having zero couplings to the cavity? The dark mode $\left|C_2\right\rangle$ and the bright mode $\left|C_1\right\rangle$ are well separated if $\delta=2Ng^2/\Delta_{ac}\gg\kappa'=\kappa+2Ng^2\gamma/\Delta_{ac}^2$, or equivalently $\sqrt{NC/2}\gg1$, where the collective cooperativity $NC=4N g^2 / \kappa \gamma$. At $\delta=0$, the dark mode is resonantly excited, while the off-resonant bright mode is only weakly generated, resulting in slightly higher power in the $\hat{a}_{+}$ traveling wave than in the $\hat{a}_{-}$ traveling wave. Therefore, the intracavity field is a partial standing wave with the field amplitudes at the positions of the atoms being minimal but not zero, as depicted in Fig.~\ref{FIG:standing_wave}(a). The atoms are weakly coupled to the cavity, providing backscattering from the $\hat{a}_{+}$ mode to the $\hat{a}_{-}$ mode while introducing very low photon loss into free space.

This nearly lossless photon transfer is analogous to the EIT, with the cavity $|+\rangle$ and $|-\rangle$ modes being two ground states and multiple atomic excited states as shown in Fig.~\ref{FIG:level}. The destructive interference of different transition pathways from $|+\rangle$ and $|-\rangle$ to $|E_1\rangle$,...,$|E_N\rangle$ leads to the formation of the dark state $\left|C_1\right\rangle$ with coupling $\mathcal{G}_2=0$ to $|A_2\rangle$, inhibiting the excitation of the atomic states and leading to zero free-space scattering.

This phenomenon is similar to the normal mode splitting caused by coupling between two traveling waves through the backscattering of atoms confined in optical lattices and nanomembranes interacting with ring cavities~\cite{klinner2006normal,yilmaz2017optomechanical} and has been applied to detect nanoparticles with use of a high-Q microresonator~\cite{zhu2010chip,peng2014parity}.

\section{Tunable photon routing with low photon loss}\label{sec:tunable_output}
\begin{figure}[!htp]
 \centering\includegraphics[width=1.0\columnwidth]{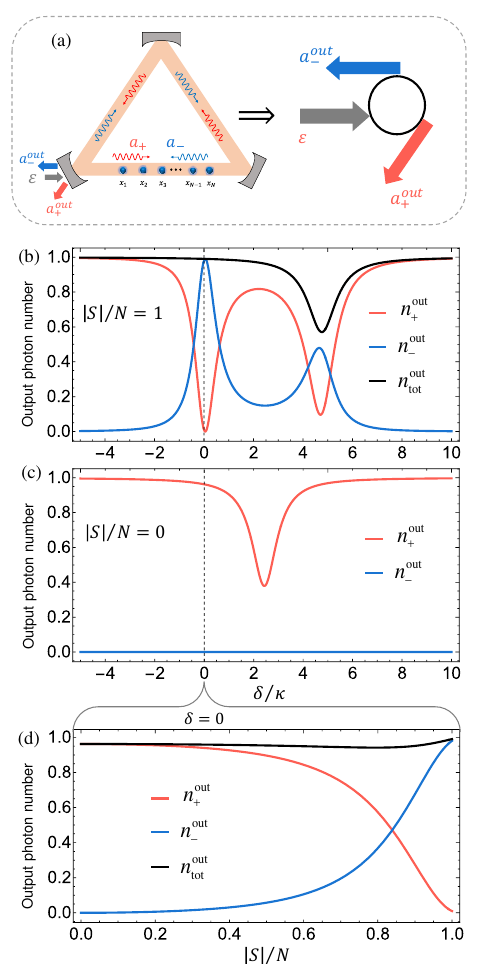}
 \caption{\label{FIG:output} (a) Realization of tunable photon routing by the ring cavity coupled to the configurable atomic chain. The normalized photon number of the cavity output modes $\hat{a}_{{+}}^{\text{out}}$, $\hat{a}_{{-}}^{\text{out}}$ in two different directions with respect to $\delta$ in the case of (b) $|\mathcal{S}|/N=1$ and (c) $|\mathcal{S}|=0$. When the drive light is resonant with the cavity, $\delta=0$, the output light is predominantly in the $\hat{a}_{-}^{\text{out}}$ direction such that $n_{-}^{\text {out}} \approx 1$ when $|\mathcal{S}|/N=1$ and is predominantly in the $\hat{a}_{+}^{\text{out}}$ direction such that $n_{+}^{\text {out}} \approx 1$ when $|\mathcal{S}|=0$. (d) Normalized output photon number as a function of the atomic structure factor $|\mathcal{S}|$ at $\delta=0$. The power ratio between the two output modes can be adjusted by variation of $|\mathcal{S}|$, while the normalized total output power (indicated by the black line) remains close to unity. This demonstrates tunable photon routing with almost no photon loss. The other parameters are $\kappa=0.1\gamma$, $g=0.5\gamma$, $\Delta_{ac}=10\gamma$, and $N=10$.
}
\end{figure}

Inspired by the lossless photon transfer between the two cavity traveling waves $\hat{a}_{+}$ and $\hat{a}_{-}$ originating from the dark mode, we next demonstrate tunable low-loss photon routing by varying the configuration of the atomic chain. The ring cavity naturally supports two output modes $\hat{a}_{+}^{\text {out}}$ and $\hat{a}_{-}^{\text{out}}$ in different directions that can be used for photon routing, as shown in Fig.~\ref{FIG:output}(a). 

Using the standard input-output relations~\cite{walls2008input}
\begin{equation} \label{eq:input-output}
\hat{a}_{\pm}^{\mathrm{in}}+\hat{a}_{\pm}^{\text {out}}=\sqrt{\kappa_{\mathrm{in}}} \hat{a}_{\pm},
\end{equation}
we obtain the steady-state output fields by substitution of Eq.~\eqref{eq:steady_state_solution} for $\hat{a}_{\pm}$ into Eq.~\eqref{eq:input-output}: 
\begin{subequations} \label{eq:output}
\begin{align}
\left\langle\hat{a}_{+}^{\text {out}}\right\rangle_{\text{ss}} &= \frac{\tilde{\Delta}\left(\tilde{\Delta} \cdot \tilde{\delta}-g^2 N\right)  \varepsilon\kappa_{\text {in }}}{\left(\tilde{\Delta} \cdot \tilde{\delta}-g^2 N\right)^2-g^4|\mathcal{S}|^2}+i \varepsilon, \\
\left\langle\hat{a}_{-}^{\text {out}}\right\rangle_{\text{ss}}&=\frac{\tilde{\Delta} g^2 \mathcal{S} \varepsilon \kappa_{\text {in }}}{\left(\tilde{\Delta} \cdot \tilde{\delta}-g^2 N\right)^2-g^4|\mathcal{S}|^2},
\end{align}
\end{subequations}
where we assume that the drive field is injected through one of the cavity mirrors and copropagates with the cavity mode $\hat{a}_{+}$, such that $\left\langle\hat{a}_{+}^{\mathrm{in}}\right\rangle=-i \varepsilon$ and $\left\langle\hat{a}_{-}^{\text {in}}\right\rangle=0$.

In Fig.~\ref{FIG:output}(b) and (c), we present the cavity output photon number normalized to the input photon number denoted as $n_{\pm}^{\text {out}}=\left|\left\langle\hat{a}_{\pm}^{\text {out}}\right\rangle\right|^2 /\left|\left\langle\hat{a}_{+}^{\text {in}}\right\rangle\right|^2$. The dependence of $n_{\pm}^{\text {out}}$ on the input-light detuning $\delta$ is examined for two distinct atomic spatial configurations, $|\mathcal{S}|/N=1$ and $|\mathcal{S}|=0$, under the condition of a large atom-cavity detuning. When $|\mathcal{S}|/N=1$, the output $n_{-}^{\text {out}}$ exhibits two peaks at the eigenfrequencies of the atom-cavity system. Particularly, when $\delta=0$, the output fields predominantly consist of the transmitted light generated from the backscattering of the atomic chain such that $n_{-}^{\text {out}} \approx 1$ and $n_{+}^{\text {out}} \approx 0$. As discussed in Section.~\ref{sec:spectrum}, in this scenario, the intracavity field is an equal superposition of $|{+}\rangle$ and $|{-}\rangle$, and the vanishing of $n_{+}^{\text {out}}$ is due to the destructive interference between the reflected light and the transmitted light in the $\hat{a}_{+}^{\text {out}}$ direction at the input-mirror interface.

However, when $|\mathcal{S}|=0$ and $\delta =0$, the output field entirely comprises the light reflected from the mirror surface such that $n_{+}^{\text{out}} \approx 1$ and $n_{-}^{\text {out}} \approx 0$. In this situation, the cavity mode $\hat{a}_{-}$ remains unexcited, and the system behaves akin to a standing-wave Fabry-P\'{e}rot cavity. Since the cavity resonance frequency is shifted by $2g^2 N / \Delta_{ac}$, the input light with $\delta=0$ is far off-resonance if $\sqrt{NC/2}\gg1$, resulting in nearly complete reflection of the input light. This is consistent with Fig.~\ref{FIG:intra}(d), which shows that the intracavity photon number is almost zero in this case.

For $0<|\mathcal{S}|<N$, there are output fields in both directions $\hat{a}_{+}^{\text{out}}$ and $\hat{a}_{-}^{\text{out}}$. When $\delta=0$, the relative power of the two output modes can be adjusted by change of the configuration $|\mathcal{S}|$ as shown in Fig.~\ref{FIG:output}(d). Notably, the normalized total output photon number $n_{\mathrm{tot}}^{\mathrm{out}}=n_{+}^{\mathrm{out}}+n_{-}^{\mathrm{out}}$ is close to unity for any value of $|\mathcal{S}|$ in the dispersive regime. Consequently, a tunable photon router is realized with almost no photon loss, as depicted in Fig.~\ref{FIG:output}(a). 

To assess the performance of the photon router with tunable relative power in two output ports, we consider two crucial technical parameters in the following discussion. The first parameter is 
\begin{equation} \label{eq:nloss}
n_{\mathrm{loss}}=\max_{|\mathcal{S}|}\left(1-n_{\mathrm{tot}}^{\mathrm{out}}\right),
\end{equation}
which is the maximum photon loss over different values of $|\mathcal{S}|$ in Fig.~\ref{FIG:output}(d), resulting from the free-space scattering. 
The analytical result for the maximum photon loss Eq.~\eqref{eq:nloss} can be derived as
\begin{small}
\begin{equation}\label{eq:max_photon_loss}
\frac{1}{NC+1}+\frac{\sqrt{(1+\frac{4\Delta^{2}}{\gamma^{2}})[(NC+1)^{2}+\frac{4\Delta^{2}}{\gamma^{2}}]}}{4(NC+1)\Delta^{2}/\gamma^{2}}-\frac{\gamma^{2}}{4\Delta^{2}}
\end{equation}
\end{small}
for $NC>2$ and $\frac{2|\Delta|}{\gamma}<\beta$, and as
\begin{small}
\begin{equation}\label{eq:max_photon_loss_2}
\frac{4NC}{(NC+1)^{2}+4\Delta^{2}/\gamma^{2}}
\end{equation}
\end{small}
for other situations. Here $\beta=(NC+1) \sqrt{\frac{NC-2}{3 NC+2}}$. Note that $\Delta=\Delta_{ac}$ since the lossless tunable photon router works at $\delta=0$. 

\begin{figure}[!htp]
 \includegraphics[width=0.90\linewidth]{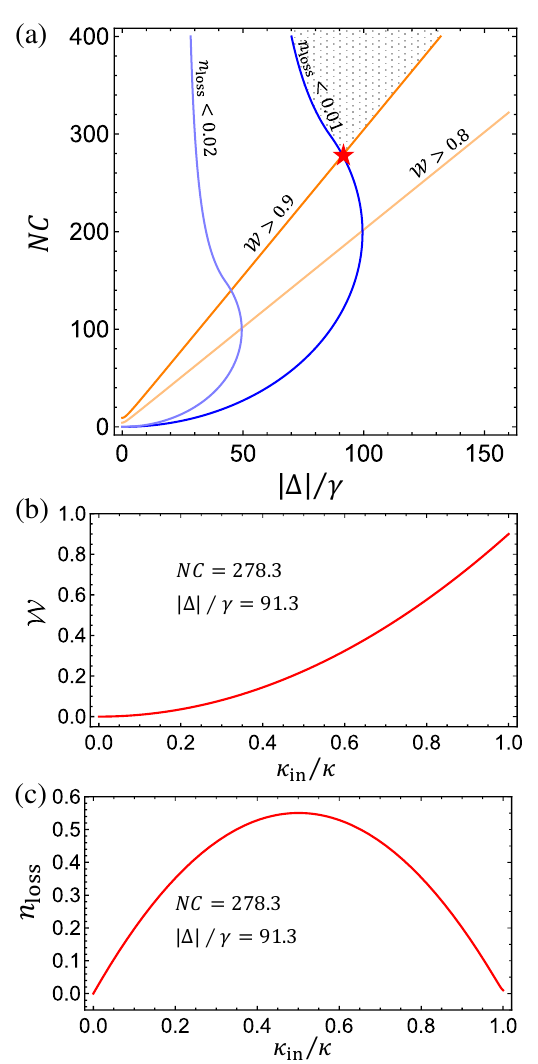}
 \caption{\label{FIG:parameter} (a) Contour plot of the maximum photon loss $n_{\mathrm{loss}}$ and the tuning range $\mathcal{W}$ with the collective cooperativity $NC$ and the normalized detuning $\Delta/\gamma$. The area to the right of the blue (light-blue) contour corresponds to $n_{\mathrm{loss}}<0.01$ ($n_{\mathrm{loss}}<0.02$). The region above the orange (light-orange) contour corresponds to $\mathcal{W}>0.9$ ($\mathcal{W}>0.8$). The shaded region above the red star indicates the overlap of $n_{\mathrm{loss}}<0.01$ and $\mathcal{W} >0.9$. (b) Tuning range $\mathcal{W}$ and (c) maximum photon loss $n_{\mathrm{loss}}$ as a function of $\kappa_{in}/\kappa$ with $NC$ and $\Delta/\gamma$ corresponding to the values at the red star in (a).}
\end{figure}

\begin{figure}
 \includegraphics[width=1.0\columnwidth]{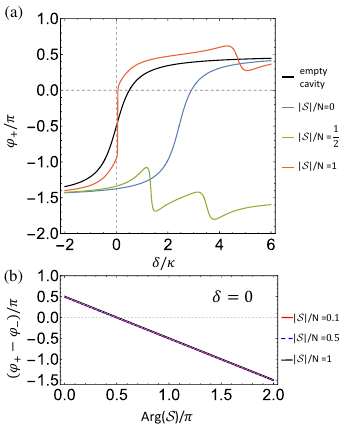}
 \caption{\label{FIG:output_phase} Phase $\varphi_{+}$ of the output mode $\hat{a}_{+}^{\text{out}}$ versus $\delta$ at selected values of $\mathcal{S}$ compared with the empty cavity. There can be a phase shift of up to $\pi$ on the cavity output when $|\mathcal{S}|/N$ is changed at a given $\delta$. (b) The relative phase $\left(\varphi_{+}-\varphi_{-}\right) / \pi$ between the modes $\hat{a}_{{+}}^{\text {out}}$ and $\hat{a}_{{-}}^{\text {out}}$ can be tuned continuously by displacement of the center of mass of the atomic chain described by the phase of $\mathcal{S}$. The phase variation is insensitive to the absolute value of $\mathcal{S}$. The other parameters are $\delta=0$, $\kappa=0.1\gamma$, $g=0.5\gamma$, $\Delta_{ac}=10\gamma$, and $N=10$.}
\end{figure}

The second parameter is the tuning range of the photon router $\mathcal{W}=\max_{|\mathcal{S}|}\left(n_{-}^{\mathrm{out}}\right)$. It can be seen from Fig.~\ref{FIG:output}(d) that $n_{-}^{\mathrm{out}}=0$ when $|\mathcal{S}|=0$, and $n_{-}^{\mathrm{out}}$ reaches the maximum value when $|\mathcal{S}|/N=1$. $\mathcal{W}$ is given by
\begin{equation} \label{eq:tune_range}
\mathcal{W} =\frac{4N^2C^2}{(2NC+1)^2+4{{\Delta ^2}/{\gamma ^2}}}. 
\end{equation}

Figure.~\ref{FIG:parameter}(a) illustrates the maximum photon loss $n_{\mathrm{loss}}$ and the tuning range $\mathcal{W}$ as a function of the collective cooperativity $NC$ and the atom-cavity detuning $|\Delta|/\gamma$. The maximum loss $n_{\mathrm{loss}}$ decreases with increasing $|\Delta|$ for fixed $NC$. This is because a large detuning reduces the scattering of photons into free space. Given a fixed $|\Delta|$, the tuning range $\mathcal{W}$ increases with $NC$ because a higher collective cooperativity generates a larger number of photons in the backward $\hat{a}_{-}$ mode. Tunable photon routing with the tuning range $\mathcal{W} \geq 0.9$ and the maximum photon loss $n_{\mathrm{loss}} \leq 0.01$ is realized in the shaded area in Fig.~\ref{FIG:parameter}(a), where $NC > 278$. This means that with a single-atom cooperativity $C \sim 30$ that is practically achievable in current experiments, an atomic chain containing as few as ten atoms can serve as an optical router. Note that the maximum photon loss $n_{\mathrm{loss}}$ converges to $\left( \sqrt{1+4{{\Delta ^2}/{\gamma ^2}}}-1 \right) /\left( 4{{\Delta ^2}/{\gamma ^2}} \right)$ when $NC \gg 1$, which indicates the existence of a minimum detuning $\left|\Delta/\gamma\right|_{\mathrm{min}} \approx 1/2n_{\text{loss}}$ if $n_{\text{loss}}\ll1$.

So far we have assumed that only the input mirror of the cavity is partially transmitting and that all the other mirrors are perfectly reflecting. In reality the nonzero transmission of the other mirrors as well as the mirror loss ($\kappa > \kappa_{\mathrm{in}}$) results in a reduction in the tuning range $\mathcal{W}$, as demonstrated in Fig.~\ref{FIG:parameter}(b). This reduction occurs due to the cavity field leakage through the other mirrors and loss. The reflected field and the transmitted field in the $\hat{a}_{{+}}^{\text {out}}$ direction at the input mirror can no longer interfere fully destructively, and $n_{+}^{\text {out}}$ is not zero. Therefore, the photon-routing range increases with the ratio of the input-mirror transmission rate to the cavity total decay rate $\kappa_{\mathrm{in}}/\kappa$. The photon leakage through other mirrors also leads to photon loss. The maximum photon loss $n_{\mathrm{loss}}$ is plotted as a function of $\kappa_{\mathrm{in}}/\kappa$ in Fig.~\ref{FIG:parameter}(c). The minimal loss $n_{\mathrm{loss}} \rightarrow 0$ when $\kappa_{\mathrm{in}}/\kappa  \rightarrow 0$ because no light can be coupled into the system.

Moreover, a tunable optical phase shift of the output fields can also be realized by a change of the atomic structure factor. As shown in Fig.~\ref{FIG:output_phase}(a), the phase $\varphi_{+}$ of the output mode $\hat{a}_{{+}}^{\text {out}}$ relative to the input mode $\hat{a}_{{+}}^{\text {in}}$ depends on the absolute value $|\mathcal{S}|$. There can be a phase shift of about $\pi$ on the cavity output $\hat{a}_{{+}}^{\text {out}}$ when $|\mathcal{S}|/N$ is changed at a given $\delta$. For $|\mathcal{S}|/N=1$, the phase $\varphi_{+}$ changes abruptly at the dark-mode resonance and smoothly near the bright mode. Similarly, for $|\mathcal{S}|/N=1/2$, there are two phase crossovers near the two shifted cavity resonances. For $|\mathcal{S}|=0$, the phase curve is akin to that of a Fabry-P\'{e}rot cavity. In addition, we can also compute the phase $\varphi_{-}$ of the output mode $\hat{a}_{{-}}^{\text {out}}$. The relative phase $\varphi_{+}-\varphi_{-}$ between the output modes $\hat{a}_{{+}}^{\text {out}}$ and $\hat{a}_{{-}}^{\text {out}}$ can be continuously tuned by displacement of the center of mass of the atomic chain, indicated by the phase of $\mathcal{S}$ as shown in Fig.~\ref{FIG:output_phase}(b). This tunable phase shift may have potential applications in realizing robust atom-photon quantum gates~\cite{reiserer2014quantum,reiserer2015cavity,reiserer2022colloquium}. 

In summary, by tuning the structure factor $\mathcal{S}$, photons can be routed between the two output modes $\hat{a}_{{+}}^{\text {out}}$ and $\hat{a}_{{-}}^{\text {out}}$, and the relative phase between the two modes $\varphi_{+}-\varphi_{-}$ can be shifted over an arbitrary range. This is because the structure factor $\mathcal{S}$ offers two degrees of freedom that can be independently controlled in experiments. One degree of freedom is the relative structure of the array described by $|\mathcal{S}|$, determined by the relative distances between the atoms. By toggling $|\mathcal{S}|/N$ between $1$ and $0$, photons are routed or switched between the two output directions. The second degree of freedom is the overall phase of $\mathcal{S}$, given by the center of mass of the atom array. By displacing the atom array and changing its center of mass—thus altering the phase of $\mathcal{S}$—the phase of the output photons can be continuously shifted in multiples of $2\pi$, enabling large phase shifts. Therefore, independent control over the relative structure of the array $|\mathcal{S}|/N$ and the center of mass of the array (the phase of $\mathcal{S}$) allows the realization of both photon routing or switching and large optical phase shifts, which is crucial in optical communication networks and quantum information processing.

\section{Experimental feasibility}\label{sec:experiment}
\begin{figure}
 \includegraphics[width=0.95\columnwidth]{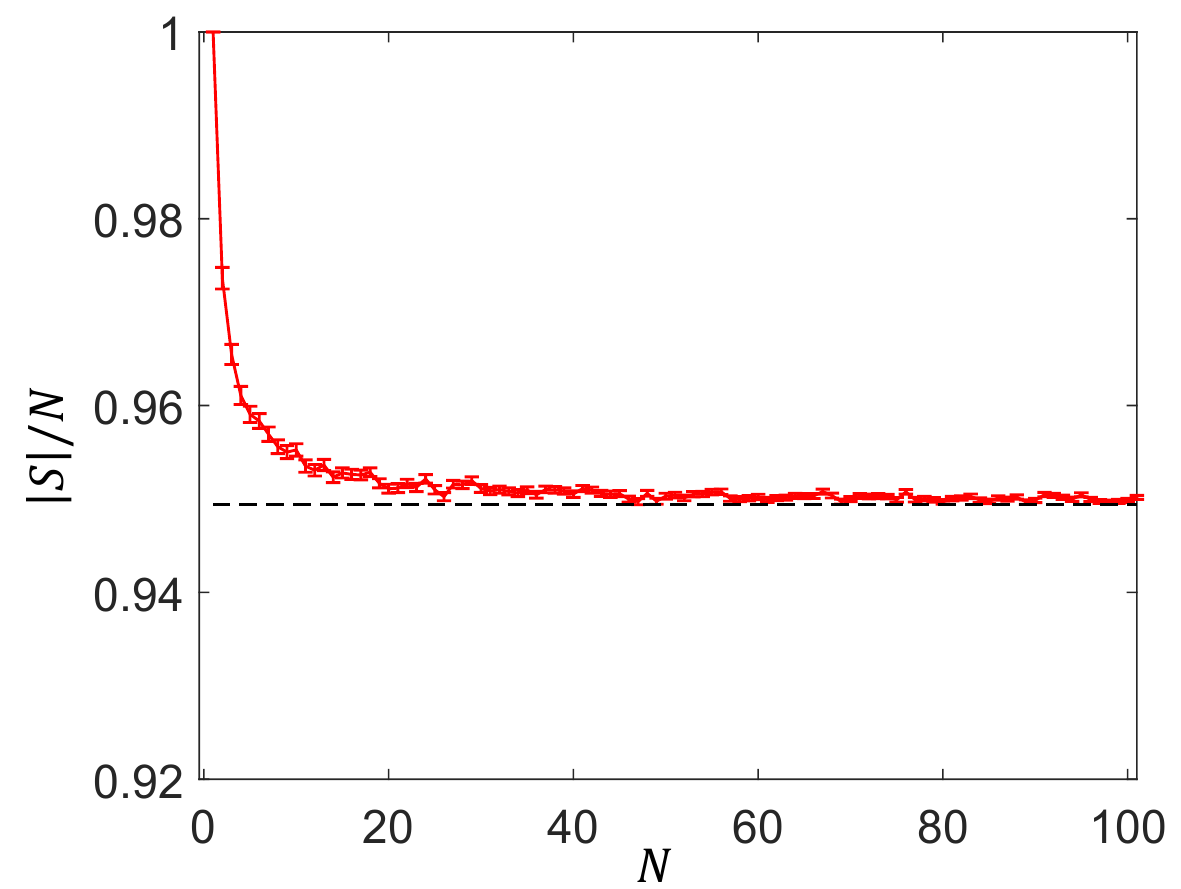}
 \caption{\label{FIG:spatial_uncertian} Numerical calculation of the reduction of the structure factor $|\mathcal{S}|/N$ resulting from atomic position uncertainty at finite temperature. The atom-array trap frequency is set to $\omega_r/2\pi = 160$kHz, and the temperature $T$ is approximately $2.5 \mu$K, leading to a standard deviation $\sigma$ of the atom position of 20 nm. The dashed line is the analytical result for $|\mathcal{S}|/N=\exp(-2 k^2 \sigma^2)$ valid for large atom number $N$.}
\end{figure}

In this section, we discuss the experimental realization of the setup.  The one-dimensional configurable atomic chain can be realized by the trapping of atoms in programmable optical tweezer arrays~\cite{endres2016atom, anderegg2019optical}. We evaluate the effect of the atomic position uncertainty on the structure factor $\mathcal{S}$ and the photon-routing tuning range $\mathcal{W}$. We take rubidium atoms as an example, in a tightly focused optical tweezer with typical beam waist $w=0.8\mu$m and the trapping depth $U/h=36$MHz, and radial trapping frequency $\omega_r=\sqrt{4U/mw^2}\simeq2\pi\times160$ kHz. With the standard Raman-sideband cooling, atoms can be cooled to temperature $T \sim 2.5 \mu$K, corresponding to approximately $95\%$ population occupation in the trap vibrational ground state. At this temperature, the standard deviation of the atom position $\sigma$ is 20 nm. The structure factor is given by $
\mathcal{S}=\sum_{i=1}^{N}\exp\left[2ik(x_i+\sigma_{x,i})\right]$, where $x_i$ is the trap center position and $\sigma_{x,i}$ is the position deviation of the $i$th atom. When $x_i$ is an integer multiple of $\lambda/2$ such that ideally $|\mathcal{S}|/N=1$, we randomly sample $\sigma_{x,i}$ with standard deviation $\sigma =20$ nm and numerically calculate $|\mathcal{S}|/N$ with the rubidium wavelength $\lambda=780$nm. As shown in Fig.~\ref{FIG:spatial_uncertian}, $|\mathcal{S}|/N$ decreases to approximately $0.95$ when $N$ is large. This is consistent with the analytical result $|\mathcal{S}|/N=\exp(-2 k^2 \sigma^2)$ for large $N$. This reduction of $|\mathcal{S}|/N =0.95 $ has a very negligible effect on the emergence of the cavity bright and dark modes, changing the interaction strength, the resonance frequencies, and the linewidths by only $5\%$. For the photon-routing performance, the slight reduction of $|\mathcal{S}|/N =0.95 $ does not change the maximum photon loss $n_{\mathrm{loss}}$, but reduces the tuning range to $\mathcal{W}=0.86$ with the parameter values in Fig.~\ref{FIG:output}(d). 

\section{Conclusions}\label{sec:conclusion}
In conclusion, we have shown that the collective atom-light interactions can be tuned by a configurable atomic chain coupled to a ring cavity. Unlike a Fabry-P\'{e}rot cavity with fixed standing-wave intracavity fields, the field pattern inside a ring cavity is determined by the spatial structure of the atomic chain. Therefore, the spectrum of the atom-cavity system is governed by the structure factor $\mathcal{S}$. When $|\mathcal{S}|=0$, the two traveling waves of the ring cavity are not coupled, resulting in a spectrum similar to that of a single-mode Fabry–P\'{e}rot cavity. However, when $|\mathcal{S}|\neq 0$, the atoms couple the two traveling waves through backscattering. Driving of the cavity forward mode $\hat{a}_{+}$ also induces excitation in the backward mode $\hat{a}_{-}$. Especially when the atoms are distributed with separations equal to integer multiples of the half-wavelength such that $|\mathcal{S}|/N=1$, the system supports a dark mode with the standing-wave nodes aligned with the atoms. This dark mode has no frequency shift or broadening since it is decoupled from the atoms. Remarkably, such a configuration of $|\mathcal{S}|/N=1$ enables lossless photon transfer from one traveling-wave mode to the other, similarly to the EIT. Furthermore, we investigate the cavity output fields and find that the photons can be routed in two cavity outgoing directions by changing the structure of the atomic chain. The tuning range $\mathcal{W}> 0.9$ and the maximum photon loss $n_{\mathrm{loss}}< 0.01$ can be reached when the collective cooperativity $NC\gtrsim 278$, which can be readily realized in experiments with an atomic chain consisting of just a few atoms. Moreover, a large optical phase shift of multiples of $2\pi$ on the cavity output can be achieved by displacement of the atomic chain. As a result, the system of atom arrays coupled to a ring cavity has significant potential for practical applications in quantum information processing and optical quantum computing.

Note that while the coupling of the photon direction with the atom-array structures can be used to realize photon switching and routing, we can reverse this capability and use the emitted directional photons as an effective quantum readout of the collective states of the atom array. Unlike destructive fluorescence images acquired via high-resolution microscopy, this readout is nondestructive since the cavity photons reveal only collective properties of the atomic system.

\section*{Acknowledgements}\label{sec:Acknowledgements}
This work was supported by the National Key Research and Development Program of China (Grant No. 2022YFA1405300) and the National Natural Science Foundation of China (Grant No. 12088101).


\begin{thebibliography}{41}%
\makeatletter
\providecommand \@ifxundefined [1]{%
 \@ifx{#1\undefined}
}%
\providecommand \@ifnum [1]{%
 \ifnum #1\expandafter \@firstoftwo
 \else \expandafter \@secondoftwo
 \fi
}%
\providecommand \@ifx [1]{%
 \ifx #1\expandafter \@firstoftwo
 \else \expandafter \@secondoftwo
 \fi
}%
\providecommand \natexlab [1]{#1}%
\providecommand \enquote  [1]{``#1''}%
\providecommand \bibnamefont  [1]{#1}%
\providecommand \bibfnamefont [1]{#1}%
\providecommand \citenamefont [1]{#1}%
\providecommand \href@noop [0]{\@secondoftwo}%
\providecommand \href [0]{\begingroup \@sanitize@url \@href}%
\providecommand \@href[1]{\@@startlink{#1}\@@href}%
\providecommand \@@href[1]{\endgroup#1\@@endlink}%
\providecommand \@sanitize@url [0]{\catcode `\\12\catcode `\$12\catcode
  `\&12\catcode `\#12\catcode `\^12\catcode `\_12\catcode `\%12\relax}%
\providecommand \@@startlink[1]{}%
\providecommand \@@endlink[0]{}%
\providecommand \url  [0]{\begingroup\@sanitize@url \@url }%
\providecommand \@url [1]{\endgroup\@href {#1}{\urlprefix }}%
\providecommand \urlprefix  [0]{URL }%
\providecommand \Eprint [0]{\href }%
\providecommand \doibase [0]{https://doi.org/}%
\providecommand \selectlanguage [0]{\@gobble}%
\providecommand \bibinfo  [0]{\@secondoftwo}%
\providecommand \bibfield  [0]{\@secondoftwo}%
\providecommand \translation [1]{[#1]}%
\providecommand \BibitemOpen [0]{}%
\providecommand \bibitemStop [0]{}%
\providecommand \bibitemNoStop [0]{.\EOS\space}%
\providecommand \EOS [0]{\spacefactor3000\relax}%
\providecommand \BibitemShut  [1]{\csname bibitem#1\endcsname}%
\let\auto@bib@innerbib\@empty
\bibitem [{\citenamefont {Haroche}\ and\ \citenamefont
  {Raimond}(2006)}]{haroche2006exploring}%
  \BibitemOpen
  \bibfield  {author} {\bibinfo {author} {\bibfnamefont {S.}~\bibnamefont
  {Haroche}}\ and\ \bibinfo {author} {\bibfnamefont {J.-M.}\ \bibnamefont
  {Raimond}},\ }\href@noop {} {\emph {\bibinfo {title} {Exploring the quantum:
  atoms, cavities, and photons}}}\ (\bibinfo  {publisher} {Oxford university
  press, Oxford},\ \bibinfo {year} {2006})\BibitemShut {NoStop}%
\bibitem [{\citenamefont {Reiserer}\ and\ \citenamefont
  {Rempe}(2015)}]{reiserer2015cavity}%
  \BibitemOpen
  \bibfield  {author} {\bibinfo {author} {\bibfnamefont {A.}~\bibnamefont
  {Reiserer}}\ and\ \bibinfo {author} {\bibfnamefont {G.}~\bibnamefont
  {Rempe}},\ }\bibfield  {title} {\bibinfo {title} {Cavity-based quantum
  networks with single atoms and optical photons},\ }\href
  {https://doi.org/10.1103/RevModPhys.87.1379} {\bibfield  {journal} {\bibinfo
  {journal} {Rev. Mod. Phys.}\ }\textbf {\bibinfo {volume} {87}},\ \bibinfo
  {pages} {1379} (\bibinfo {year} {2015})}\BibitemShut {NoStop}%
\bibitem [{\citenamefont {Reiserer}(2022)}]{reiserer2022colloquium}%
  \BibitemOpen
  \bibfield  {author} {\bibinfo {author} {\bibfnamefont {A.}~\bibnamefont
  {Reiserer}},\ }\bibfield  {title} {\bibinfo {title} {Colloquium:
  Cavity-enhanced quantum network nodes},\ }\href
  {https://doi.org/10.1103/RevModPhys.94.041003} {\bibfield  {journal}
  {\bibinfo  {journal} {Rev. Mod. Phys.}\ }\textbf {\bibinfo {volume} {94}},\
  \bibinfo {pages} {041003} (\bibinfo {year} {2022})}\BibitemShut {NoStop}%
\bibitem [{\citenamefont {Ritsch}\ \emph {et~al.}(2013)\citenamefont {Ritsch},
  \citenamefont {Domokos}, \citenamefont {Brennecke},\ and\ \citenamefont
  {Esslinger}}]{ritsch2013cold}%
  \BibitemOpen
  \bibfield  {author} {\bibinfo {author} {\bibfnamefont {H.}~\bibnamefont
  {Ritsch}}, \bibinfo {author} {\bibfnamefont {P.}~\bibnamefont {Domokos}},
  \bibinfo {author} {\bibfnamefont {F.}~\bibnamefont {Brennecke}},\ and\
  \bibinfo {author} {\bibfnamefont {T.}~\bibnamefont {Esslinger}},\ }\bibfield
  {title} {\bibinfo {title} {Cold atoms in cavity-generated dynamical optical
  potentials},\ }\href {https://doi.org/10.1103/RevModPhys.85.553} {\bibfield
  {journal} {\bibinfo  {journal} {Rev. Mod. Phys.}\ }\textbf {\bibinfo {volume}
  {85}},\ \bibinfo {pages} {553} (\bibinfo {year} {2013})}\BibitemShut
  {NoStop}%
\bibitem [{\citenamefont {Chang}\ \emph {et~al.}(2018)\citenamefont {Chang},
  \citenamefont {Douglas}, \citenamefont {Gonz\'alez-Tudela}, \citenamefont
  {Hung},\ and\ \citenamefont {Kimble}}]{chang2018colloquium}%
  \BibitemOpen
  \bibfield  {author} {\bibinfo {author} {\bibfnamefont {D.~E.}\ \bibnamefont
  {Chang}}, \bibinfo {author} {\bibfnamefont {J.~S.}\ \bibnamefont {Douglas}},
  \bibinfo {author} {\bibfnamefont {A.}~\bibnamefont {Gonz\'alez-Tudela}},
  \bibinfo {author} {\bibfnamefont {C.-L.}\ \bibnamefont {Hung}},\ and\
  \bibinfo {author} {\bibfnamefont {H.~J.}\ \bibnamefont {Kimble}},\ }\bibfield
   {title} {\bibinfo {title} {Colloquium: Quantum matter built from nanoscopic
  lattices of atoms and photons},\ }\href
  {https://doi.org/10.1103/RevModPhys.90.031002} {\bibfield  {journal}
  {\bibinfo  {journal} {Rev. Mod. Phys.}\ }\textbf {\bibinfo {volume} {90}},\
  \bibinfo {pages} {031002} (\bibinfo {year} {2018})}\BibitemShut {NoStop}%
\bibitem [{\citenamefont {Mivehvar}\ \emph {et~al.}(2021)\citenamefont
  {Mivehvar}, \citenamefont {Piazza}, \citenamefont {Donner},\ and\
  \citenamefont {Ritsch}}]{mivehvar2021cavity}%
  \BibitemOpen
  \bibfield  {author} {\bibinfo {author} {\bibfnamefont {F.}~\bibnamefont
  {Mivehvar}}, \bibinfo {author} {\bibfnamefont {F.}~\bibnamefont {Piazza}},
  \bibinfo {author} {\bibfnamefont {T.}~\bibnamefont {Donner}},\ and\ \bibinfo
  {author} {\bibfnamefont {H.}~\bibnamefont {Ritsch}},\ }\bibfield  {title}
  {\bibinfo {title} {Cavity qed with quantum gases: new paradigms in many-body
  physics},\ }\href {https://doi.org/10.1080/00018732.2021.1969727} {\bibfield
  {journal} {\bibinfo  {journal} {Adv. Phys}\ }\textbf {\bibinfo
  {volume} {70}},\ \bibinfo {pages} {1} (\bibinfo {year} {2021})}\BibitemShut
  {NoStop}%
\bibitem [{\citenamefont {Norcia}\ \emph {et~al.}(2018)\citenamefont {Norcia},
  \citenamefont {Lewis-Swan}, \citenamefont {Cline}, \citenamefont {Zhu},
  \citenamefont {Rey},\ and\ \citenamefont {Thompson}}]{norcia2018cavity}%
  \BibitemOpen
  \bibfield  {author} {\bibinfo {author} {\bibfnamefont {M.~A.}\ \bibnamefont
  {Norcia}}, \bibinfo {author} {\bibfnamefont {R.~J.}\ \bibnamefont
  {Lewis-Swan}}, \bibinfo {author} {\bibfnamefont {J.~R.}\ \bibnamefont
  {Cline}}, \bibinfo {author} {\bibfnamefont {B.}~\bibnamefont {Zhu}}, \bibinfo
  {author} {\bibfnamefont {A.~M.}\ \bibnamefont {Rey}},\ and\ \bibinfo {author}
  {\bibfnamefont {J.~K.}\ \bibnamefont {Thompson}},\ }\bibfield  {title}
  {\bibinfo {title} {Cavity-mediated collective spin-exchange interactions in a
  strontium superradiant laser},\ }\href
  {https://doi.org/10.1126/science.aar3102} {\bibfield  {journal} {\bibinfo
  {journal} {Science}\ }\textbf {\bibinfo {volume} {361}},\ \bibinfo {pages}
  {259} (\bibinfo {year} {2018})}\BibitemShut {NoStop}%
\bibitem [{\citenamefont {Periwal}\ \emph {et~al.}(2021)\citenamefont
  {Periwal}, \citenamefont {Cooper}, \citenamefont {Kunkel}, \citenamefont
  {Wienand}, \citenamefont {Davis},\ and\ \citenamefont
  {Schleier-Smith}}]{periwal2021programmable}%
  \BibitemOpen
  \bibfield  {author} {\bibinfo {author} {\bibfnamefont {A.}~\bibnamefont
  {Periwal}}, \bibinfo {author} {\bibfnamefont {E.~S.}\ \bibnamefont {Cooper}},
  \bibinfo {author} {\bibfnamefont {P.}~\bibnamefont {Kunkel}}, \bibinfo
  {author} {\bibfnamefont {J.~F.}\ \bibnamefont {Wienand}}, \bibinfo {author}
  {\bibfnamefont {E.~J.}\ \bibnamefont {Davis}},\ and\ \bibinfo {author}
  {\bibfnamefont {M.}~\bibnamefont {Schleier-Smith}},\ }\bibfield  {title}
  {\bibinfo {title} {Programmable interactions and emergent geometry in an
  array of atom clouds},\ }\href {https://doi.org/10.1038/s41586-021-04156-0}
  {\bibfield  {journal} {\bibinfo  {journal} {Nature}\ }\textbf {\bibinfo
  {volume} {600}},\ \bibinfo {pages} {630} (\bibinfo {year}
  {2021})}\BibitemShut {NoStop}%
\bibitem [{\citenamefont {Vaneecloo}\ \emph {et~al.}(2022)\citenamefont
  {Vaneecloo}, \citenamefont {Garcia},\ and\ \citenamefont
  {Ourjoumtsev}}]{vaneecloo2022}%
  \BibitemOpen
  \bibfield  {author} {\bibinfo {author} {\bibfnamefont {J.}~\bibnamefont
  {Vaneecloo}}, \bibinfo {author} {\bibfnamefont {S.}~\bibnamefont {Garcia}},\
  and\ \bibinfo {author} {\bibfnamefont {A.}~\bibnamefont {Ourjoumtsev}},\
  }\bibfield  {title} {\bibinfo {title} {Intracavity rydberg superatom for
  optical quantum engineering: Coherent control, single-shot detection, and
  optical $\ensuremath{\pi}$ phase shift},\ }\href
  {https://doi.org/10.1103/PhysRevX.12.021034} {\bibfield  {journal} {\bibinfo
  {journal} {Phys. Rev. X}\ }\textbf {\bibinfo {volume} {12}},\ \bibinfo
  {pages} {021034} (\bibinfo {year} {2022})}\BibitemShut {NoStop}%
\bibitem [{\citenamefont {Stolz}\ \emph {et~al.}(2022)\citenamefont {Stolz},
  \citenamefont {Hegels}, \citenamefont {Winter}, \citenamefont {R\"ohr},
  \citenamefont {Hsiao}, \citenamefont {Husel}, \citenamefont {Rempe},\ and\
  \citenamefont {D\"urr}}]{stolz2022}%
  \BibitemOpen
  \bibfield  {author} {\bibinfo {author} {\bibfnamefont {T.}~\bibnamefont
  {Stolz}}, \bibinfo {author} {\bibfnamefont {H.}~\bibnamefont {Hegels}},
  \bibinfo {author} {\bibfnamefont {M.}~\bibnamefont {Winter}}, \bibinfo
  {author} {\bibfnamefont {B.}~\bibnamefont {R\"ohr}}, \bibinfo {author}
  {\bibfnamefont {Y.-F.}\ \bibnamefont {Hsiao}}, \bibinfo {author}
  {\bibfnamefont {L.}~\bibnamefont {Husel}}, \bibinfo {author} {\bibfnamefont
  {G.}~\bibnamefont {Rempe}},\ and\ \bibinfo {author} {\bibfnamefont
  {S.}~\bibnamefont {D\"urr}},\ }\bibfield  {title} {\bibinfo {title}
  {Quantum-logic gate between two optical photons with an average efficiency
  above 40\%},\ }\href {https://doi.org/10.1103/PhysRevX.12.021035} {\bibfield
  {journal} {\bibinfo  {journal} {Phys. Rev. X}\ }\textbf {\bibinfo {volume}
  {12}},\ \bibinfo {pages} {021035} (\bibinfo {year} {2022})}\BibitemShut
  {NoStop}%
\bibitem [{\citenamefont {Leroux}\ \emph {et~al.}(2010)\citenamefont {Leroux},
  \citenamefont {Schleier-Smith},\ and\ \citenamefont
  {Vuleti\ifmmode~\acute{c}\else \'{c}\fi{}}}]{leroux2010implementation}%
  \BibitemOpen
  \bibfield  {author} {\bibinfo {author} {\bibfnamefont {I.~D.}\ \bibnamefont
  {Leroux}}, \bibinfo {author} {\bibfnamefont {M.~H.}\ \bibnamefont
  {Schleier-Smith}},\ and\ \bibinfo {author} {\bibfnamefont {V.}~\bibnamefont
  {Vuleti\ifmmode~\acute{c}\else \'{c}\fi{}}},\ }\bibfield  {title} {\bibinfo
  {title} {Implementation of cavity squeezing of a collective atomic spin},\
  }\href {https://doi.org/10.1103/PhysRevLett.104.073602} {\bibfield  {journal}
  {\bibinfo  {journal} {Phys. Rev. Lett.}\ }\textbf {\bibinfo {volume} {104}},\
  \bibinfo {pages} {073602} (\bibinfo {year} {2010})}\BibitemShut {NoStop}%
\bibitem [{\citenamefont {Haas}\ \emph {et~al.}(2014)\citenamefont {Haas},
  \citenamefont {Volz}, \citenamefont {Gehr}, \citenamefont {Reichel},\ and\
  \citenamefont {Est{\`e}ve}}]{haas2014entangled}%
  \BibitemOpen
  \bibfield  {author} {\bibinfo {author} {\bibfnamefont {F.}~\bibnamefont
  {Haas}}, \bibinfo {author} {\bibfnamefont {J.}~\bibnamefont {Volz}}, \bibinfo
  {author} {\bibfnamefont {R.}~\bibnamefont {Gehr}}, \bibinfo {author}
  {\bibfnamefont {J.}~\bibnamefont {Reichel}},\ and\ \bibinfo {author}
  {\bibfnamefont {J.}~\bibnamefont {Est{\`e}ve}},\ }\bibfield  {title}
  {\bibinfo {title} {Entangled states of more than 40 atoms in an optical fiber
  cavity},\ }\href {https://doi.org/10.1126/science.1248905} {\bibfield
  {journal} {\bibinfo  {journal} {Science}\ }\textbf {\bibinfo {volume}
  {344}},\ \bibinfo {pages} {180} (\bibinfo {year} {2014})}\BibitemShut
  {NoStop}%
\bibitem [{\citenamefont {Welte}\ \emph {et~al.}(2017)\citenamefont {Welte},
  \citenamefont {Hacker}, \citenamefont {Daiss}, \citenamefont {Ritter},\ and\
  \citenamefont {Rempe}}]{welte2017cavity}%
  \BibitemOpen
  \bibfield  {author} {\bibinfo {author} {\bibfnamefont {S.}~\bibnamefont
  {Welte}}, \bibinfo {author} {\bibfnamefont {B.}~\bibnamefont {Hacker}},
  \bibinfo {author} {\bibfnamefont {S.}~\bibnamefont {Daiss}}, \bibinfo
  {author} {\bibfnamefont {S.}~\bibnamefont {Ritter}},\ and\ \bibinfo {author}
  {\bibfnamefont {G.}~\bibnamefont {Rempe}},\ }\bibfield  {title} {\bibinfo
  {title} {Cavity carving of atomic bell states},\ }\href
  {https://doi.org/10.1103/PhysRevLett.118.210503} {\bibfield  {journal}
  {\bibinfo  {journal} {Phys. Rev. Lett.}\ }\textbf {\bibinfo {volume} {118}},\
  \bibinfo {pages} {210503} (\bibinfo {year} {2017})}\BibitemShut {NoStop}%
\bibitem [{\citenamefont {Hosten}\ \emph {et~al.}(2016)\citenamefont {Hosten},
  \citenamefont {Engelsen}, \citenamefont {Krishnakumar},\ and\ \citenamefont
  {Kasevich}}]{hosten2016measurement}%
  \BibitemOpen
  \bibfield  {author} {\bibinfo {author} {\bibfnamefont {O.}~\bibnamefont
  {Hosten}}, \bibinfo {author} {\bibfnamefont {N.~J.}\ \bibnamefont
  {Engelsen}}, \bibinfo {author} {\bibfnamefont {R.}~\bibnamefont
  {Krishnakumar}},\ and\ \bibinfo {author} {\bibfnamefont {M.~A.}\ \bibnamefont
  {Kasevich}},\ }\bibfield  {title} {\bibinfo {title} {Measurement noise 100
  times lower than the quantum-projection limit using entangled atoms},\ }\href
  {https://doi.org/10.1038/nature16176} {\bibfield  {journal} {\bibinfo
  {journal} {Nature}\ }\textbf {\bibinfo {volume} {529}},\ \bibinfo {pages}
  {505} (\bibinfo {year} {2016})}\BibitemShut {NoStop}%
\bibitem [{\citenamefont {Pedrozo-Pe{\~n}afiel}\ \emph
  {et~al.}(2020)\citenamefont {Pedrozo-Pe{\~n}afiel}, \citenamefont {Colombo},
  \citenamefont {Shu}, \citenamefont {Adiyatullin}, \citenamefont {Li},
  \citenamefont {Mendez}, \citenamefont {Braverman}, \citenamefont {Kawasaki},
  \citenamefont {Akamatsu}, \citenamefont {Xiao} \emph
  {et~al.}}]{pedrozo2020entanglement}%
  \BibitemOpen
  \bibfield  {author} {\bibinfo {author} {\bibfnamefont {E.}~\bibnamefont
  {Pedrozo-Pe{\~n}afiel}}, \bibinfo {author} {\bibfnamefont {S.}~\bibnamefont
  {Colombo}}, \bibinfo {author} {\bibfnamefont {C.}~\bibnamefont {Shu}},
  \bibinfo {author} {\bibfnamefont {A.~F.}\ \bibnamefont {Adiyatullin}},
  \bibinfo {author} {\bibfnamefont {Z.}~\bibnamefont {Li}}, \bibinfo {author}
  {\bibfnamefont {E.}~\bibnamefont {Mendez}}, \bibinfo {author} {\bibfnamefont
  {B.}~\bibnamefont {Braverman}}, \bibinfo {author} {\bibfnamefont
  {A.}~\bibnamefont {Kawasaki}}, \bibinfo {author} {\bibfnamefont
  {D.}~\bibnamefont {Akamatsu}}, \bibinfo {author} {\bibfnamefont
  {Y.}~\bibnamefont {Xiao}}, \emph {et~al.},\ }\bibfield  {title} {\bibinfo
  {title} {Entanglement on an optical atomic-clock transition},\ }\href@noop {}
  {\bibfield  {journal} {\bibinfo  {journal} {Nature}\ }\textbf {\bibinfo
  {volume} {588}},\ \bibinfo {pages} {414} (\bibinfo {year}
  {2020})}\BibitemShut {NoStop}%
\bibitem [{\citenamefont {Greve}\ \emph {et~al.}(2022)\citenamefont {Greve},
  \citenamefont {Luo}, \citenamefont {Wu},\ and\ \citenamefont
  {Thompson}}]{greve2022entanglement}%
  \BibitemOpen
  \bibfield  {author} {\bibinfo {author} {\bibfnamefont {G.~P.}\ \bibnamefont
  {Greve}}, \bibinfo {author} {\bibfnamefont {C.}~\bibnamefont {Luo}}, \bibinfo
  {author} {\bibfnamefont {B.}~\bibnamefont {Wu}},\ and\ \bibinfo {author}
  {\bibfnamefont {J.~K.}\ \bibnamefont {Thompson}},\ }\bibfield  {title}
  {\bibinfo {title} {Entanglement-enhanced matter-wave interferometry in a
  high-finesse cavity},\ }\href@noop {} {\bibfield  {journal} {\bibinfo
  {journal} {Nature}\ }\textbf {\bibinfo {volume} {610}},\ \bibinfo {pages}
  {472} (\bibinfo {year} {2022})}\BibitemShut {NoStop}%
\bibitem [{\citenamefont {Reimann}\ \emph {et~al.}(2015)\citenamefont
  {Reimann}, \citenamefont {Alt}, \citenamefont {Kampschulte}, \citenamefont
  {Macha}, \citenamefont {Ratschbacher}, \citenamefont {Thau}, \citenamefont
  {Yoon},\ and\ \citenamefont {Meschede}}]{reimann2015cavity}%
  \BibitemOpen
  \bibfield  {author} {\bibinfo {author} {\bibfnamefont {R.}~\bibnamefont
  {Reimann}}, \bibinfo {author} {\bibfnamefont {W.}~\bibnamefont {Alt}},
  \bibinfo {author} {\bibfnamefont {T.}~\bibnamefont {Kampschulte}}, \bibinfo
  {author} {\bibfnamefont {T.}~\bibnamefont {Macha}}, \bibinfo {author}
  {\bibfnamefont {L.}~\bibnamefont {Ratschbacher}}, \bibinfo {author}
  {\bibfnamefont {N.}~\bibnamefont {Thau}}, \bibinfo {author} {\bibfnamefont
  {S.}~\bibnamefont {Yoon}},\ and\ \bibinfo {author} {\bibfnamefont
  {D.}~\bibnamefont {Meschede}},\ }\bibfield  {title} {\bibinfo {title}
  {Cavity-modified collective rayleigh scattering of two atoms},\ }\href
  {https://doi.org/10.1103/PhysRevLett.114.023601} {\bibfield  {journal}
  {\bibinfo  {journal} {Phys. Rev. Lett.}\ }\textbf {\bibinfo {volume} {114}},\
  \bibinfo {pages} {023601} (\bibinfo {year} {2015})}\BibitemShut {NoStop}%
\bibitem [{\citenamefont {Deist}\ \emph
  {et~al.}(2022{\natexlab{a}})\citenamefont {Deist}, \citenamefont {Gerber},
  \citenamefont {Lu}, \citenamefont {Zeiher},\ and\ \citenamefont
  {Stamper-Kurn}}]{deist2022superresolution}%
  \BibitemOpen
  \bibfield  {author} {\bibinfo {author} {\bibfnamefont {E.}~\bibnamefont
  {Deist}}, \bibinfo {author} {\bibfnamefont {J.~A.}\ \bibnamefont {Gerber}},
  \bibinfo {author} {\bibfnamefont {Y.-H.}\ \bibnamefont {Lu}}, \bibinfo
  {author} {\bibfnamefont {J.}~\bibnamefont {Zeiher}},\ and\ \bibinfo {author}
  {\bibfnamefont {D.~M.}\ \bibnamefont {Stamper-Kurn}},\ }\bibfield  {title}
  {\bibinfo {title} {Superresolution microscopy of optical fields using
  tweezer-trapped single atoms},\ }\href
  {https://doi.org/10.1103/PhysRevLett.128.083201} {\bibfield  {journal}
  {\bibinfo  {journal} {Phys. Rev. Lett.}\ }\textbf {\bibinfo {volume} {128}},\
  \bibinfo {pages} {083201} (\bibinfo {year} {2022}{\natexlab{a}})}\BibitemShut
  {NoStop}%
\bibitem [{\citenamefont {Deist}\ \emph
  {et~al.}(2022{\natexlab{b}})\citenamefont {Deist}, \citenamefont {Lu},
  \citenamefont {Ho}, \citenamefont {Pasha}, \citenamefont {Zeiher},
  \citenamefont {Yan},\ and\ \citenamefont {Stamper-Kurn}}]{deist2022mid}%
  \BibitemOpen
  \bibfield  {author} {\bibinfo {author} {\bibfnamefont {E.}~\bibnamefont
  {Deist}}, \bibinfo {author} {\bibfnamefont {Y.-H.}\ \bibnamefont {Lu}},
  \bibinfo {author} {\bibfnamefont {J.}~\bibnamefont {Ho}}, \bibinfo {author}
  {\bibfnamefont {M.~K.}\ \bibnamefont {Pasha}}, \bibinfo {author}
  {\bibfnamefont {J.}~\bibnamefont {Zeiher}}, \bibinfo {author} {\bibfnamefont
  {Z.}~\bibnamefont {Yan}},\ and\ \bibinfo {author} {\bibfnamefont {D.~M.}\
  \bibnamefont {Stamper-Kurn}},\ }\bibfield  {title} {\bibinfo {title}
  {Mid-circuit cavity measurement in a neutral atom array},\ }\href
  {https://doi.org/10.1103/PhysRevLett.129.203602} {\bibfield  {journal}
  {\bibinfo  {journal} {Phys. Rev. Lett.}\ }\textbf {\bibinfo {volume} {129}},\
  \bibinfo {pages} {203602} (\bibinfo {year} {2022}{\natexlab{b}})}\BibitemShut
  {NoStop}%
\bibitem [{\citenamefont {Liu}\ \emph {et~al.}(2023)\citenamefont {Liu},
  \citenamefont {Wang}, \citenamefont {Yang}, \citenamefont {Wang},
  \citenamefont {Fan}, \citenamefont {Guan}, \citenamefont {Li}, \citenamefont
  {Zhang},\ and\ \citenamefont {Zhang}}]{liu2023realization}%
  \BibitemOpen
  \bibfield  {author} {\bibinfo {author} {\bibfnamefont {Y.}~\bibnamefont
  {Liu}}, \bibinfo {author} {\bibfnamefont {Z.}~\bibnamefont {Wang}}, \bibinfo
  {author} {\bibfnamefont {P.}~\bibnamefont {Yang}}, \bibinfo {author}
  {\bibfnamefont {Q.}~\bibnamefont {Wang}}, \bibinfo {author} {\bibfnamefont
  {Q.}~\bibnamefont {Fan}}, \bibinfo {author} {\bibfnamefont {S.}~\bibnamefont
  {Guan}}, \bibinfo {author} {\bibfnamefont {G.}~\bibnamefont {Li}}, \bibinfo
  {author} {\bibfnamefont {P.}~\bibnamefont {Zhang}},\ and\ \bibinfo {author}
  {\bibfnamefont {T.}~\bibnamefont {Zhang}},\ }\bibfield  {title} {\bibinfo
  {title} {Realization of strong coupling between deterministic single-atom
  arrays and a high-finesse miniature optical cavity},\ }\href
  {https://doi.org/10.1103/PhysRevLett.130.173601} {\bibfield  {journal}
  {\bibinfo  {journal} {Phys. Rev. Lett.}\ }\textbf {\bibinfo {volume} {130}},\
  \bibinfo {pages} {173601} (\bibinfo {year} {2023})}\BibitemShut {NoStop}%
\bibitem [{\citenamefont {Nagorny}\ \emph {et~al.}(2003)\citenamefont
  {Nagorny}, \citenamefont {Els\"asser},\ and\ \citenamefont
  {Hemmerich}}]{nagorny2003collective}%
  \BibitemOpen
  \bibfield  {author} {\bibinfo {author} {\bibfnamefont {B.}~\bibnamefont
  {Nagorny}}, \bibinfo {author} {\bibfnamefont {T.}~\bibnamefont
  {Els\"asser}},\ and\ \bibinfo {author} {\bibfnamefont {A.}~\bibnamefont
  {Hemmerich}},\ }\bibfield  {title} {\bibinfo {title} {Collective atomic
  motion in an optical lattice formed inside a high finesse cavity},\ }\href
  {https://doi.org/10.1103/PhysRevLett.91.153003} {\bibfield  {journal}
  {\bibinfo  {journal} {Phys. Rev. Lett.}\ }\textbf {\bibinfo {volume} {91}},\
  \bibinfo {pages} {153003} (\bibinfo {year} {2003})}\BibitemShut {NoStop}%
\bibitem [{\citenamefont {Jia}\ \emph {et~al.}(2018)\citenamefont {Jia},
  \citenamefont {Schine}, \citenamefont {Georgakopoulos}, \citenamefont {Ryou},
  \citenamefont {Clark}, \citenamefont {Sommer},\ and\ \citenamefont
  {Simon}}]{jia2018strongly}%
  \BibitemOpen
  \bibfield  {author} {\bibinfo {author} {\bibfnamefont {N.}~\bibnamefont
  {Jia}}, \bibinfo {author} {\bibfnamefont {N.}~\bibnamefont {Schine}},
  \bibinfo {author} {\bibfnamefont {A.}~\bibnamefont {Georgakopoulos}},
  \bibinfo {author} {\bibfnamefont {A.}~\bibnamefont {Ryou}}, \bibinfo {author}
  {\bibfnamefont {L.~W.}\ \bibnamefont {Clark}}, \bibinfo {author}
  {\bibfnamefont {A.}~\bibnamefont {Sommer}},\ and\ \bibinfo {author}
  {\bibfnamefont {J.}~\bibnamefont {Simon}},\ }\bibfield  {title} {\bibinfo
  {title} {A strongly interacting polaritonic quantum dot},\ }\href
  {https://www.nature.com/articles/s41567-018-0071-6} {\bibfield  {journal}
  {\bibinfo  {journal} {Nat. Phys}\ }\textbf {\bibinfo {volume} {14}},\
  \bibinfo {pages} {550} (\bibinfo {year} {2018})}\BibitemShut {NoStop}%
\bibitem [{\citenamefont {Ostermann}\ \emph {et~al.}(2020)\citenamefont
  {Ostermann}, \citenamefont {Niedenzu},\ and\ \citenamefont
  {Ritsch}}]{ostermann2020unraveling}%
  \BibitemOpen
  \bibfield  {author} {\bibinfo {author} {\bibfnamefont {S.}~\bibnamefont
  {Ostermann}}, \bibinfo {author} {\bibfnamefont {W.}~\bibnamefont
  {Niedenzu}},\ and\ \bibinfo {author} {\bibfnamefont {H.}~\bibnamefont
  {Ritsch}},\ }\bibfield  {title} {\bibinfo {title} {Unraveling the quantum
  nature of atomic self-ordering in a ring cavity},\ }\href
  {https://doi.org/10.1103/PhysRevLett.124.033601} {\bibfield  {journal}
  {\bibinfo  {journal} {Phys. Rev. Lett.}\ }\textbf {\bibinfo {volume} {124}},\
  \bibinfo {pages} {033601} (\bibinfo {year} {2020})}\BibitemShut {NoStop}%
\bibitem [{\citenamefont {Schuster}\ \emph {et~al.}(2020)\citenamefont
  {Schuster}, \citenamefont {Wolf}, \citenamefont {Ostermann}, \citenamefont
  {Slama},\ and\ \citenamefont {Zimmermann}}]{schuster2020supersolid}%
  \BibitemOpen
  \bibfield  {author} {\bibinfo {author} {\bibfnamefont {S.~C.}\ \bibnamefont
  {Schuster}}, \bibinfo {author} {\bibfnamefont {P.}~\bibnamefont {Wolf}},
  \bibinfo {author} {\bibfnamefont {S.}~\bibnamefont {Ostermann}}, \bibinfo
  {author} {\bibfnamefont {S.}~\bibnamefont {Slama}},\ and\ \bibinfo {author}
  {\bibfnamefont {C.}~\bibnamefont {Zimmermann}},\ }\bibfield  {title}
  {\bibinfo {title} {Supersolid properties of a bose-einstein condensate in a
  ring resonator},\ }\href {https://doi.org/10.1103/PhysRevLett.124.143602}
  {\bibfield  {journal} {\bibinfo  {journal} {Phys. Rev. Lett.}\ }\textbf
  {\bibinfo {volume} {124}},\ \bibinfo {pages} {143602} (\bibinfo {year}
  {2020})}\BibitemShut {NoStop}%
\bibitem [{\citenamefont {Mivehvar}\ \emph {et~al.}(2018)\citenamefont
  {Mivehvar}, \citenamefont {Ostermann}, \citenamefont {Piazza},\ and\
  \citenamefont {Ritsch}}]{mivehvar2018driven}%
  \BibitemOpen
  \bibfield  {author} {\bibinfo {author} {\bibfnamefont {F.}~\bibnamefont
  {Mivehvar}}, \bibinfo {author} {\bibfnamefont {S.}~\bibnamefont {Ostermann}},
  \bibinfo {author} {\bibfnamefont {F.}~\bibnamefont {Piazza}},\ and\ \bibinfo
  {author} {\bibfnamefont {H.}~\bibnamefont {Ritsch}},\ }\bibfield  {title}
  {\bibinfo {title} {Driven-dissipative supersolid in a ring cavity},\ }\href
  {https://doi.org/10.1103/PhysRevLett.120.123601} {\bibfield  {journal}
  {\bibinfo  {journal} {Phys. Rev. Lett.}\ }\textbf {\bibinfo {volume} {120}},\
  \bibinfo {pages} {123601} (\bibinfo {year} {2018})}\BibitemShut {NoStop}%
\bibitem [{\citenamefont {Reiserer}\ \emph {et~al.}(2014)\citenamefont
  {Reiserer}, \citenamefont {Kalb}, \citenamefont {Rempe},\ and\ \citenamefont
  {Ritter}}]{reiserer2014quantum}%
  \BibitemOpen
  \bibfield  {author} {\bibinfo {author} {\bibfnamefont {A.}~\bibnamefont
  {Reiserer}}, \bibinfo {author} {\bibfnamefont {N.}~\bibnamefont {Kalb}},
  \bibinfo {author} {\bibfnamefont {G.}~\bibnamefont {Rempe}},\ and\ \bibinfo
  {author} {\bibfnamefont {S.}~\bibnamefont {Ritter}},\ }\bibfield  {title}
  {\bibinfo {title} {A quantum gate between a flying optical photon and a
  single trapped atom},\ }\href {https://doi.org/10.1038/nature13177}
  {\bibfield  {journal} {\bibinfo  {journal} {Nature}\ }\textbf {\bibinfo
  {volume} {508}},\ \bibinfo {pages} {237} (\bibinfo {year}
  {2014})}\BibitemShut {NoStop}%
\bibitem [{\citenamefont {Shomroni}\ \emph {et~al.}(2014)\citenamefont
  {Shomroni}, \citenamefont {Rosenblum}, \citenamefont {Lovsky}, \citenamefont
  {Bechler}, \citenamefont {Guendelman},\ and\ \citenamefont
  {Dayan}}]{shomroni2014all}%
  \BibitemOpen
  \bibfield  {author} {\bibinfo {author} {\bibfnamefont {I.}~\bibnamefont
  {Shomroni}}, \bibinfo {author} {\bibfnamefont {S.}~\bibnamefont {Rosenblum}},
  \bibinfo {author} {\bibfnamefont {Y.}~\bibnamefont {Lovsky}}, \bibinfo
  {author} {\bibfnamefont {O.}~\bibnamefont {Bechler}}, \bibinfo {author}
  {\bibfnamefont {G.}~\bibnamefont {Guendelman}},\ and\ \bibinfo {author}
  {\bibfnamefont {B.}~\bibnamefont {Dayan}},\ }\bibfield  {title} {\bibinfo
  {title} {All-optical routing of single photons by a one-atom switch
  controlled by a single photon},\ }\href
  {https://www.science.org/doi/10.1126/science.1254699} {\bibfield  {journal}
  {\bibinfo  {journal} {Science}\ }\textbf {\bibinfo {volume} {345}},\ \bibinfo
  {pages} {903} (\bibinfo {year} {2014})}\BibitemShut {NoStop}%
\bibitem [{\citenamefont {Scheucher}\ \emph {et~al.}(2016)\citenamefont
  {Scheucher}, \citenamefont {Hilico}, \citenamefont {Will}, \citenamefont
  {Volz},\ and\ \citenamefont {Rauschenbeutel}}]{scheucher2016quantum}%
  \BibitemOpen
  \bibfield  {author} {\bibinfo {author} {\bibfnamefont {M.}~\bibnamefont
  {Scheucher}}, \bibinfo {author} {\bibfnamefont {A.}~\bibnamefont {Hilico}},
  \bibinfo {author} {\bibfnamefont {E.}~\bibnamefont {Will}}, \bibinfo {author}
  {\bibfnamefont {J.}~\bibnamefont {Volz}},\ and\ \bibinfo {author}
  {\bibfnamefont {A.}~\bibnamefont {Rauschenbeutel}},\ }\bibfield  {title}
  {\bibinfo {title} {Quantum optical circulator controlled by a single chirally
  coupled atom},\ }\href {https://www.science.org/doi/10.1126/science.aaj2118}
  {\bibfield  {journal} {\bibinfo  {journal} {Science}\ }\textbf {\bibinfo
  {volume} {354}},\ \bibinfo {pages} {1577} (\bibinfo {year}
  {2016})}\BibitemShut {NoStop}%
\bibitem [{\citenamefont {Xia}\ and\ \citenamefont
  {Twamley}(2013)}]{xia2013all}%
  \BibitemOpen
  \bibfield  {author} {\bibinfo {author} {\bibfnamefont {K.}~\bibnamefont
  {Xia}}\ and\ \bibinfo {author} {\bibfnamefont {J.}~\bibnamefont {Twamley}},\
  }\bibfield  {title} {\bibinfo {title} {All-optical switching and router via
  the direct quantum control of coupling between cavity modes},\ }\href
  {https://doi.org/10.1103/PhysRevX.3.031013} {\bibfield  {journal} {\bibinfo
  {journal} {Phys. Rev. X}\ }\textbf {\bibinfo {volume} {3}},\ \bibinfo {pages}
  {031013} (\bibinfo {year} {2013})}\BibitemShut {NoStop}%
\bibitem [{\citenamefont {O'brien}(2007)}]{o2007optical}%
  \BibitemOpen
  \bibfield  {author} {\bibinfo {author} {\bibfnamefont {J.~L.}\ \bibnamefont
  {O'brien}},\ }\bibfield  {title} {\bibinfo {title} {Optical quantum
  computing},\ }\href {https://www.science.org/doi/10.1126/science.1142892}
  {\bibfield  {journal} {\bibinfo  {journal} {Science}\ }\textbf {\bibinfo
  {volume} {318}},\ \bibinfo {pages} {1567} (\bibinfo {year}
  {2007})}\BibitemShut {NoStop}%
\bibitem [{\citenamefont {Kok}\ \emph {et~al.}(2007)\citenamefont {Kok},
  \citenamefont {Munro}, \citenamefont {Nemoto}, \citenamefont {Ralph},
  \citenamefont {Dowling},\ and\ \citenamefont {Milburn}}]{kok2007linear}%
  \BibitemOpen
  \bibfield  {author} {\bibinfo {author} {\bibfnamefont {P.}~\bibnamefont
  {Kok}}, \bibinfo {author} {\bibfnamefont {W.~J.}\ \bibnamefont {Munro}},
  \bibinfo {author} {\bibfnamefont {K.}~\bibnamefont {Nemoto}}, \bibinfo
  {author} {\bibfnamefont {T.~C.}\ \bibnamefont {Ralph}}, \bibinfo {author}
  {\bibfnamefont {J.~P.}\ \bibnamefont {Dowling}},\ and\ \bibinfo {author}
  {\bibfnamefont {G.~J.}\ \bibnamefont {Milburn}},\ }\bibfield  {title}
  {\bibinfo {title} {Linear optical quantum computing with photonic qubits},\
  }\href {https://journals.aps.org/rmp/abstract/10.1103/RevModPhys.79.135}
  {\bibfield  {journal} {\bibinfo  {journal} {Reviews of modern physics}\
  }\textbf {\bibinfo {volume} {79}},\ \bibinfo {pages} {135} (\bibinfo {year}
  {2007})}\BibitemShut {NoStop}%
\bibitem [{\citenamefont {Tavis}\ and\ \citenamefont
  {Cummings}(1968)}]{tavis1968exact}%
  \BibitemOpen
  \bibfield  {author} {\bibinfo {author} {\bibfnamefont {M.}~\bibnamefont
  {Tavis}}\ and\ \bibinfo {author} {\bibfnamefont {F.~W.}\ \bibnamefont
  {Cummings}},\ }\bibfield  {title} {\bibinfo {title} {Exact solution for an
  $n$-molecule---radiation-field hamiltonian},\ }\href
  {https://doi.org/10.1103/PhysRev.170.379} {\bibfield  {journal} {\bibinfo
  {journal} {Phys. Rev.}\ }\textbf {\bibinfo {volume} {170}},\ \bibinfo {pages}
  {379} (\bibinfo {year} {1968})}\BibitemShut {NoStop}%
\bibitem [{\citenamefont {Wickenbrock}\ \emph {et~al.}(2013)\citenamefont
  {Wickenbrock}, \citenamefont {Hemmerling}, \citenamefont {Robb},
  \citenamefont {Emary},\ and\ \citenamefont
  {Renzoni}}]{wickenbrock2013collective}%
  \BibitemOpen
  \bibfield  {author} {\bibinfo {author} {\bibfnamefont {A.}~\bibnamefont
  {Wickenbrock}}, \bibinfo {author} {\bibfnamefont {M.}~\bibnamefont
  {Hemmerling}}, \bibinfo {author} {\bibfnamefont {G.~R.~M.}\ \bibnamefont
  {Robb}}, \bibinfo {author} {\bibfnamefont {C.}~\bibnamefont {Emary}},\ and\
  \bibinfo {author} {\bibfnamefont {F.}~\bibnamefont {Renzoni}},\ }\bibfield
  {title} {\bibinfo {title} {Collective strong coupling in multimode cavity
  qed},\ }\href {https://doi.org/10.1103/PhysRevA.87.043817} {\bibfield
  {journal} {\bibinfo  {journal} {Phys. Rev. A}\ }\textbf {\bibinfo {volume}
  {87}},\ \bibinfo {pages} {043817} (\bibinfo {year} {2013})}\BibitemShut
  {NoStop}%
\bibitem [{\citenamefont {Emary}(2013)}]{emary2013dark}%
  \BibitemOpen
  \bibfield  {author} {\bibinfo {author} {\bibfnamefont {C.}~\bibnamefont
  {Emary}},\ }\bibfield  {title} {\bibinfo {title} {Dark-states in multi-mode
  multi-atom jaynes–cummings systems},\ }\href
  {https://doi.org/10.1088/0953-4075/46/22/224008} {\bibfield  {journal}
  {\bibinfo  {journal} {J. Phys. B: At., Mol. Opt.
  Phys.}\ }\textbf {\bibinfo {volume} {46}},\ \bibinfo {pages} {224008}
  (\bibinfo {year} {2013})}\BibitemShut {NoStop}%
\bibitem [{\citenamefont {Klinner}\ \emph {et~al.}(2006)\citenamefont
  {Klinner}, \citenamefont {Lindholdt}, \citenamefont {Nagorny},\ and\
  \citenamefont {Hemmerich}}]{klinner2006normal}%
  \BibitemOpen
  \bibfield  {author} {\bibinfo {author} {\bibfnamefont {J.}~\bibnamefont
  {Klinner}}, \bibinfo {author} {\bibfnamefont {M.}~\bibnamefont {Lindholdt}},
  \bibinfo {author} {\bibfnamefont {B.}~\bibnamefont {Nagorny}},\ and\ \bibinfo
  {author} {\bibfnamefont {A.}~\bibnamefont {Hemmerich}},\ }\bibfield  {title}
  {\bibinfo {title} {Normal mode splitting and mechanical effects of an optical
  lattice in a ring cavity},\ }\href
  {https://doi.org/10.1103/PhysRevLett.96.023002} {\bibfield  {journal}
  {\bibinfo  {journal} {Phys. Rev. Lett.}\ }\textbf {\bibinfo {volume} {96}},\
  \bibinfo {pages} {023002} (\bibinfo {year} {2006})}\BibitemShut {NoStop}%
\bibitem [{\citenamefont {Yilmaz}\ \emph {et~al.}(2017)\citenamefont {Yilmaz},
  \citenamefont {Schuster}, \citenamefont {Wolf}, \citenamefont {Schmidt},
  \citenamefont {Eisele}, \citenamefont {Zimmermann},\ and\ \citenamefont
  {Slama}}]{yilmaz2017optomechanical}%
  \BibitemOpen
  \bibfield  {author} {\bibinfo {author} {\bibfnamefont {A.}~\bibnamefont
  {Yilmaz}}, \bibinfo {author} {\bibfnamefont {S.}~\bibnamefont {Schuster}},
  \bibinfo {author} {\bibfnamefont {P.}~\bibnamefont {Wolf}}, \bibinfo {author}
  {\bibfnamefont {D.}~\bibnamefont {Schmidt}}, \bibinfo {author} {\bibfnamefont
  {M.}~\bibnamefont {Eisele}}, \bibinfo {author} {\bibfnamefont
  {C.}~\bibnamefont {Zimmermann}},\ and\ \bibinfo {author} {\bibfnamefont
  {S.}~\bibnamefont {Slama}},\ }\bibfield  {title} {\bibinfo {title}
  {Optomechanical damping of a nanomembrane inside an optical ring cavity},\
  }\href {https://doi.org/10.1088/1367-2630/aa55ee} {\bibfield  {journal}
  {\bibinfo  {journal} {New. J. Phys}\ }\textbf {\bibinfo {volume}
  {19}},\ \bibinfo {pages} {013038} (\bibinfo {year} {2017})}\BibitemShut
  {NoStop}%
\bibitem [{\citenamefont {Zhu}\ \emph {et~al.}(2010)\citenamefont {Zhu},
  \citenamefont {Ozdemir}, \citenamefont {Xiao}, \citenamefont {Li},
  \citenamefont {He}, \citenamefont {Chen},\ and\ \citenamefont
  {Yang}}]{zhu2010chip}%
  \BibitemOpen
  \bibfield  {author} {\bibinfo {author} {\bibfnamefont {J.}~\bibnamefont
  {Zhu}}, \bibinfo {author} {\bibfnamefont {S.~K.}\ \bibnamefont {Ozdemir}},
  \bibinfo {author} {\bibfnamefont {Y.-F.}\ \bibnamefont {Xiao}}, \bibinfo
  {author} {\bibfnamefont {L.}~\bibnamefont {Li}}, \bibinfo {author}
  {\bibfnamefont {L.}~\bibnamefont {He}}, \bibinfo {author} {\bibfnamefont
  {D.-R.}\ \bibnamefont {Chen}},\ and\ \bibinfo {author} {\bibfnamefont
  {L.}~\bibnamefont {Yang}},\ }\bibfield  {title} {\bibinfo {title} {On-chip
  single nanoparticle detection and sizing by mode splitting in an ultrahigh-q
  microresonator},\ }\href
  {https://doi.org/https://doi.org/10.1038/nphoton.2009.237} {\bibfield
  {journal} {\bibinfo  {journal} {Nat. photonics}\ }\textbf {\bibinfo
  {volume} {4}},\ \bibinfo {pages} {46} (\bibinfo {year} {2010})}\BibitemShut
  {NoStop}%
\bibitem [{\citenamefont {Peng}\ \emph {et~al.}(2014)\citenamefont {Peng},
  \citenamefont {{\"O}zdemir}, \citenamefont {Lei}, \citenamefont {Monifi},
  \citenamefont {Gianfreda}, \citenamefont {Long}, \citenamefont {Fan},
  \citenamefont {Nori}, \citenamefont {Bender},\ and\ \citenamefont
  {Yang}}]{peng2014parity}%
  \BibitemOpen
  \bibfield  {author} {\bibinfo {author} {\bibfnamefont {B.}~\bibnamefont
  {Peng}}, \bibinfo {author} {\bibfnamefont {{\c{S}}.~K.}\ \bibnamefont
  {{\"O}zdemir}}, \bibinfo {author} {\bibfnamefont {F.}~\bibnamefont {Lei}},
  \bibinfo {author} {\bibfnamefont {F.}~\bibnamefont {Monifi}}, \bibinfo
  {author} {\bibfnamefont {M.}~\bibnamefont {Gianfreda}}, \bibinfo {author}
  {\bibfnamefont {G.~L.}\ \bibnamefont {Long}}, \bibinfo {author}
  {\bibfnamefont {S.}~\bibnamefont {Fan}}, \bibinfo {author} {\bibfnamefont
  {F.}~\bibnamefont {Nori}}, \bibinfo {author} {\bibfnamefont {C.~M.}\
  \bibnamefont {Bender}},\ and\ \bibinfo {author} {\bibfnamefont
  {L.}~\bibnamefont {Yang}},\ }\bibfield  {title} {\bibinfo {title}
  {Parity--time-symmetric whispering-gallery microcavities},\ }\href
  {https://doi.org/https://doi.org/10.1038/nphys2927} {\bibfield  {journal}
  {\bibinfo  {journal} {Nat. Phys}\ }\textbf {\bibinfo {volume} {10}},\
  \bibinfo {pages} {394} (\bibinfo {year} {2014})}\BibitemShut {NoStop}%
\bibitem [{\citenamefont {Walls}\ and\ \citenamefont
  {Milburn}(2008)}]{walls2008input}%
  \BibitemOpen
  \bibfield  {author} {\bibinfo {author} {\bibfnamefont {D.}~\bibnamefont
  {Walls}}\ and\ \bibinfo {author} {\bibfnamefont {G.~J.}\ \bibnamefont
  {Milburn}},\ }\bibfield  {title} {\bibinfo {title} {Input--output formulation
  of optical cavities},\ }in\ \href@noop {} {\emph {\bibinfo {booktitle}
  {Quantum optics}}}\ (\bibinfo  {publisher} {Springer},\ \bibinfo {year}
  {2008})\ pp.\ \bibinfo {pages} {127--141}\BibitemShut {NoStop}%
\bibitem [{\citenamefont {Endres}\ \emph {et~al.}(2016)\citenamefont {Endres},
  \citenamefont {Bernien}, \citenamefont {Keesling}, \citenamefont {Levine},
  \citenamefont {Anschuetz}, \citenamefont {Krajenbrink}, \citenamefont
  {Senko}, \citenamefont {Vuletic}, \citenamefont {Greiner},\ and\
  \citenamefont {Lukin}}]{endres2016atom}%
  \BibitemOpen
  \bibfield  {author} {\bibinfo {author} {\bibfnamefont {M.}~\bibnamefont
  {Endres}}, \bibinfo {author} {\bibfnamefont {H.}~\bibnamefont {Bernien}},
  \bibinfo {author} {\bibfnamefont {A.}~\bibnamefont {Keesling}}, \bibinfo
  {author} {\bibfnamefont {H.}~\bibnamefont {Levine}}, \bibinfo {author}
  {\bibfnamefont {E.~R.}\ \bibnamefont {Anschuetz}}, \bibinfo {author}
  {\bibfnamefont {A.}~\bibnamefont {Krajenbrink}}, \bibinfo {author}
  {\bibfnamefont {C.}~\bibnamefont {Senko}}, \bibinfo {author} {\bibfnamefont
  {V.}~\bibnamefont {Vuletic}}, \bibinfo {author} {\bibfnamefont
  {M.}~\bibnamefont {Greiner}},\ and\ \bibinfo {author} {\bibfnamefont {M.~D.}\
  \bibnamefont {Lukin}},\ }\bibfield  {title} {\bibinfo {title} {Atom-by-atom
  assembly of defect-free one-dimensional cold atom arrays},\ }\href
  {https://www.science.org/doi/full/10.1126/science.aah3752} {\bibfield
  {journal} {\bibinfo  {journal} {Science}\ }\textbf {\bibinfo {volume}
  {354}},\ \bibinfo {pages} {1024} (\bibinfo {year} {2016})}\BibitemShut
  {NoStop}%
\bibitem [{\citenamefont {Anderegg}\ \emph {et~al.}(2019)\citenamefont
  {Anderegg}, \citenamefont {Cheuk}, \citenamefont {Bao}, \citenamefont
  {Burchesky}, \citenamefont {Ketterle}, \citenamefont {Ni},\ and\
  \citenamefont {Doyle}}]{anderegg2019optical}%
  \BibitemOpen
  \bibfield  {author} {\bibinfo {author} {\bibfnamefont {L.}~\bibnamefont
  {Anderegg}}, \bibinfo {author} {\bibfnamefont {L.~W.}\ \bibnamefont {Cheuk}},
  \bibinfo {author} {\bibfnamefont {Y.}~\bibnamefont {Bao}}, \bibinfo {author}
  {\bibfnamefont {S.}~\bibnamefont {Burchesky}}, \bibinfo {author}
  {\bibfnamefont {W.}~\bibnamefont {Ketterle}}, \bibinfo {author}
  {\bibfnamefont {K.-K.}\ \bibnamefont {Ni}},\ and\ \bibinfo {author}
  {\bibfnamefont {J.~M.}\ \bibnamefont {Doyle}},\ }\bibfield  {title} {\bibinfo
  {title} {An optical tweezer array of ultracold molecules},\ }\href
  {https://www.science.org/doi/full/10.1126/science.aax1265} {\bibfield
  {journal} {\bibinfo  {journal} {Science}\ }\textbf {\bibinfo {volume}
  {365}},\ \bibinfo {pages} {1156} (\bibinfo {year} {2019})}\BibitemShut
  {NoStop}%
\end{thebibliography}

%

\end{document}